\documentclass[amsmath,amssymb,aps,pre,twocolumn,superscriptaddress]{revtex4-2} %twocolumn
\usepackage{amsmath,amsfonts}
\usepackage{soul}
\usepackage{dsfont}
\usepackage{graphicx}% Include figure files
\usepackage{bm}% bold math
\usepackage{hyperref}% add hypertext capabilities
\usepackage{url}
\usepackage[english]{babel}
\usepackage[utf8x]{inputenc}
\usepackage[T1]{fontenc}

\usepackage{xcolor}

\newcommand{\mi}{\mathrm{i}}

\def\fig#1{{Fig.~\ref{#1}}}     

\begin{document}

\title{Spontaneous axis formation through defect fusion in active nematics}
%Spontaneous axis formation by fusion of topological defects through active nematic traction
%Surface nematics on a sphere: active tractions as a mechanism for symmetry breaking}

\author{Mathieu Dedenon}
\email[Contact author: ]{mdedenon@pks.mpg.de}
\altaffiliation[Current address: ]{Max Planck Institute for the Physics of Complex Systems, N\"othnitzer Stra\ss e 38, 01187 Dresden, Germany}
\affiliation{Department of Biochemistry, University of Geneva, 1211 Geneva, Switzerland}
\affiliation{Department of Theoretical Physics, University of Geneva, 1211 Geneva, Switzerland}

\author{Claire A. Dessalles}
%\email{claire.dessalles@univ-lyon1.fr}
\affiliation{Department of Biochemistry, University of Geneva, 1211 Geneva, Switzerland}
\affiliation{Université Claude Bernard Lyon 1, CNRS, Institut Lumière Matière, UMR5306, F-69100, Villeurbanne, France}

%\author{Hélène Delanoë-Ayari}
%\email{Helene.Delanoe-Ayari@univ-lyon1.fr}
%\affiliation{Université Claude Bernard Lyon 1, CNRS, Institut Lumière Matière, UMR5306, F-69100, Villeurbanne, France}

\author{Aur\'elien Roux}
%\email{Aurelien.Roux@unige.ch}
\affiliation{Department of Biochemistry, University of Geneva, 1211 Geneva, Switzerland}

\author{Karsten Kruse}
\email[Contact author: ]{Karsten.Kruse@unige.ch}
\affiliation{Department of Biochemistry, University of Geneva, 1211 Geneva, Switzerland}
\affiliation{Department of Theoretical Physics, University of Geneva, 1211 Geneva, Switzerland}

\date{\today}
\begin{abstract}
Initially spherical embryos break symmetry to establish a body axis. Topological defects in tissue nematic order have been implicated in axis formation, but the physical mechanisms underlying defect organization in active materials remain unclear. We study an active nematic fluid on a sphere using a fully covariant hydrodynamic theory that ensures thermodynamic consistency in the passive limit. We find that, in compressible systems, active nematic traction drives a spontaneous transition from the equilibrium configuration of four separated $+1/2$ defects to two antipodal $+1$ defects through pairwise fusion. The resulting state defines an axis, with aster or vortex textures at the poles depending on the flow–nematic coupling. These results identify a physical mechanism by which active forces reorganize topological defects to establish an axis on a spherical surface.
\end{abstract}

\maketitle

%\begin{linenumbers}

\paragraph*{Introduction.}
Active fluids with orientational order provide a framework for understanding how local force generation produces collective motion and spatial organization in living matter~\cite{Prost:2015ev,Marchetti2013,Doostmohammadi.2018}. Applications range from cytoskeletal filament-motor mixtures to cell layers and tissues~\cite{Salbreux:2009fp,Sanchez2012,Saw:2017gn,Kawaguchi:2017em,Chugh2018,Copenhagen.2021}. Beyond generating flows, orientational order can organize biological form: nematic alignment accompanies the cytokinetic abscission during cell division~\cite{Fishkind1993,Turlier:2014hq}, convergent extension during gastrulation~\cite{Li.2025}, while topological defects in the supracellular actin network of \textit{Hydra} are associated with organizing centers of morphogenesis~\cite{MaroudasSacks2020,Ravichandran2025}. Understanding how active forces organize these defects is therefore a step toward connecting local mechanics to tissue-scale pattern formation.

%Fueled by the continuous consumption of chemical energy, these systems exhibit a wealth of nonequilibrium phenomena, including spontaneous flow generation, defect nucleation, and active turbulence~\cite{Marchetti2013,Doostmohammadi.2018,Alert.2021}, that have no counterpart in passive liquid crystals. Active nematic theories have successfully described \textit{in vitro} mixtures of cytoskeletal filaments and molecular motors~\cite{Sanchez2012}, the actomyosin cortex underlying the plasma membrane of animal cells~\cite{Salbreux:2009fp}, bacterial colonies~\cite{Copenhagen.2021}, and epithelial monolayers~\cite{Saw:2017gn,Kawaguchi:2017em}.
%Although active nematics have been studied predominantly on flat substrates, many biological active materials are intrinsically confined to curved surfaces, which can influence their function~\cite{Luciano.2024}. Prominent biological examples include the cortical actomyosin network~\cite{Turlier:2014hq,Chugh2018}, the supracellular actin fibers that organize the body surface of regenerating \textit{Hydra}~\cite{MaroudasSacks2020,Ravichandran2025}, and the epithelial shell of the blastula~\cite{Li.2025}. The dynamics of active nematics on curved surfaces has been studied in reconstituted systems~\cite{Keber2014,Giomi2017,Maechler.2019,Pearce2019,Dessalles2025,Guillamat2026}.

Many biological materials are intrinsically confined to curved surfaces, which can influence their function~\cite{Maechler.2019,Ehrig2019,Luciano.2024}. On closed surfaces, topology constrains the possible defect configurations. A tangent nematic field on a sphere must carry a total topological charge of $+2$~\cite{Vitelli2006}. For a passive nematic in the one-constant approximation, orientational elasticity favors four $+1/2$ defects arranged at the vertices of a tetrahedron. Biological textures can exhibit a different organization. In the freshwater polyp \textit{Hydra}, $+1$ defects in the supracellular actin orientation are associated with the mouth and foot, linking defect organization to the body axis~\cite{MaroudasSacks2020,Ravichandran2025}.
Similarly, cortical actomyosin reorganizes along the division axis towards the cytokinetic ring during cell division~\cite{Fishkind1993,Turlier:2014hq}.
How can activity reorganize the nematic texture into two integer defects instead of energetically favorable half-integer defects, and how can they be positioned at opposite poles to define a single axis?

Previous works have shown that activity profoundly modifies defect dynamics on curved surfaces. Self-propelled $+1/2$-defects can oscillate between tetrahedral and planar configurations on a sphere~\cite{Keber2014}. More generally, curvature and surface deformation couple to nematic order and defect motion~\cite{Giomi2017,Pearce2019,Ehrig2019,Vafa2022,Dessalles2025,Guillamat2026}.

In active nematics, flows are commonly driven by a nematic active stress. Alternatively, active traction provides coupling between nematic organization and force generation through interaction with a supporting substrate. Recent experiments and theory identified nonlinear active forces that account for cell accumulation around imposed $+1$ defects~\cite{Guillamat:2022,Dedenon.2023,Zhao2025}. How these forces affect defect organization when the nematic field can evolve is largely unknown.

Motivated by experiemnts that will be described elsewhere, we demonstrate that active surface traction can fundamentally change the organization of active nematics on a spherical shell. Using a covariant hydrodynamic theory, we show that they can generate large-scale surface flows that overcome the elastic repulsion between neighboring $+1/2$ defects. These flows can drive fusion of $+1/2$ defects into $+1$ defects, which lie at opposite poles. Active surface traction therefore provides a generic physical mechanism through which closed active surfaces spontaneously establish an axis.
Beyond explaining how integer defect configurations emerge, our results suggest a route by which active stress can translate topological constraints into body-axis formation during morphogenesis and regeneration.

\begin{figure}
\centering
\includegraphics[width=\linewidth]{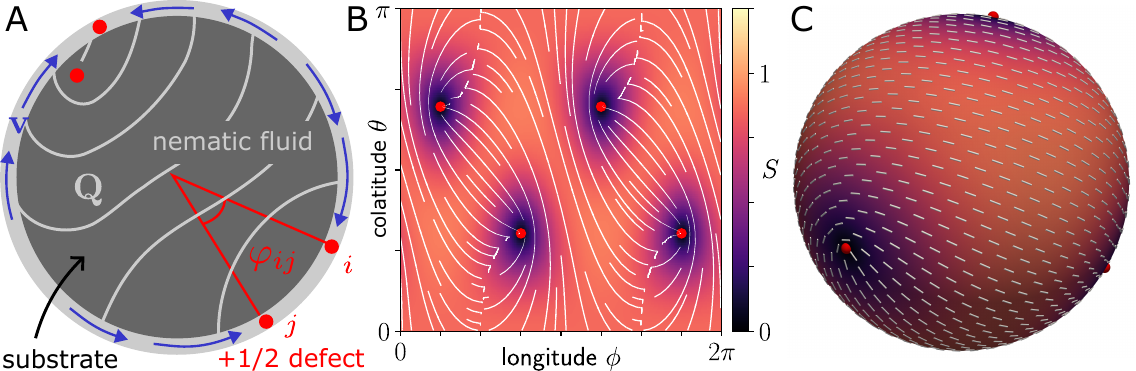}
\caption{\textit{Illustration of the system considered.} (A) Cut through a sphere to illustrate a  surfacic nematic fluid (light gray) in contact with a spherical substrate (dark gray). Orientational order ($\mathbf Q$) is also indicated on the half sphere facing the reader, the velocity field $\mathbf v$ (blue) only in the plane of the cut. Red points represent positions of $+1/2$-defects, $\varphi_{ij}$ denotes the angular distance between defects $i$ and $j$.
(B) Mercator plot of the equilibrium nematic texture with four $+1/2$-defects (red points). White lines: nematic orientation, color code: nematic order amplitude $S$.
(C) Three-dimensional view of the nematic field corresponding to panel (B). Gray lines: local nematic directors, color code: as in (B). Parameters are $\chi=6$, $\xi_s=0.5$, $\alpha=\beta=0$, $\lambda=1$.
}
\label{fig1}
\end{figure}

\paragraph*{Theory.}%of surface nematics
We consider a two-dimensional nematic fluid on a fixed sphere of radius $R$, \fig{fig1}A. Its state is specified by the tangent velocity field $\mathbf v$ and the tangent nematic tensor $\mathbf Q=S[2\hat{\mathbf n}\hat{\mathbf n}-\bm 1]$, where $S$ is the nematic order parameter, $\hat{\mathbf n}$ the unit director and $\bm 1$ is the tangential identity tensor. 

The dynamics of the nematic tensor is given by 
\begin{align}\label{eq:dtQ}
(\partial_t+\mathbf v\cdot\bm D)\mathbf Q+\bm\omega\cdot\mathbf Q-\mathbf Q\cdot\bm\omega=\frac{1}{\Gamma}\mathbf H+2\lambda\tilde{\mathbf u}.
\end{align}
Here, $\bm D$ denotes the covariant derivative, $\Gamma$ is the rotational viscosity, $\lambda$ is the flow-alignment parameter, $\bm\omega=\tfrac{1}{2}(\bm D\mathbf v-\bm D\mathbf v^T)$ is the vorticity tensor, and $\tilde{\mathbf u}=\tfrac{1}{2}(\bm D\mathbf v+\bm D\mathbf v^T)-\tfrac{1}{2}\left(\bm D\cdot\mathbf{v}\right)\bm 1$ the traceless strain-rate tensor. The molecular field $\mathbf H=KD^2\mathbf Q+\chi(1-S^2)\mathbf Q$ derives from a free energy $\mathcal F=\int\mathrm{d}Q\,f$ with 
\begin{align}
f=-\frac{\chi}{2}\mathbf Q:\mathbf Q+\frac{\chi}{8}(\mathbf Q:\mathbf Q)^2+\frac{K}{2}(\bm D\mathbf Q)^2.
\end{align}
Here, the parameter $\chi>0$ favors a nematic order $S=1$ and the orientational elasticity $K$ penalizes spatial variations of $\mathbf Q$ in the one-constant approximation.

At zero Reynolds number, the velocity field $\mathbf v$ is determined by force balance
\begin{align}\label{eq:force-cov}
\bm D\cdot\bm\sigma &=\xi_{\rm s}\mathbf v-\mathbf f_{\rm s},
\end{align}
where $\xi_s$ is the substrate friction coefficient. The stress tensor $\bm\sigma$ is
\begin{align}
\label{eq:stresstraceless}
\bm\sigma &=-P\bm 1+\tilde{\bm\sigma}^{\rm E}-2\lambda\mathbf H+2\eta\tilde{\mathbf u}+\alpha\mathbf Q,
\end{align}
with shear viscosity $\eta$ and active stress coefficient $\alpha$. In this expression, $P$ denotes the isotropic part and $\tilde{\bm\sigma}^{\rm E}$ is the traceless Ericksen stress
\begin{align}
\tilde{\bm\sigma}^{\rm E}=-K\bm D\mathbf Q:\bm D\mathbf Q+\frac{K}{2}(\bm D\mathbf Q)^2\bm 1+\mathbf H\cdot\mathbf Q-\mathbf Q\cdot\mathbf H,
\end{align}
with $(\bm D\mathbf Q:\bm D\mathbf Q)_{ij}=(D_iQ^{kl})(D_jQ_{kl})$.
%Furthermore, the parameter $\eta$ denotes the shear viscosity, $\alpha$ measures the active stress, and $\xi_{\rm s}$ quantifies friction between the fluid and the surface.
%Finally, for the stress \eqref{eq:stresstraceless}, the coupling between orientational elasticity and curvature requires an additional term to assure thermodynamic consistency in the passive case $\alpha=0$~\cite{Bouck2023}, see also App.~\ref{app:relaxation}. Alternatively, the correct equilibrium behavior can be obtained by an external force in addition to the friction between the active fluid and the surface. We opt for the latter and the corresponding force is $\mathbf{f}_{\rm s}$ in the force balance equation~\eqref{eq:force-cov}.

On a curved surface, orientational dynamics and force balance require additional curvature contributions to ensure thermodynamic consistency~\cite{Bouck2023}.
%Finally to assure thermodynamic consistency in the passive case $\alpha=0$, the coupling between orientational elasticity and curvature requires an additional term ~\cite{Bouck2023}.
This contribution would take the form of an internal stress if the theory was written in terms of an external surface derivative accounting for normal variations. Here, our choice to use only covariant derivatives implies that an external force $\mathbf{f}_{\rm s}$ must be added in \eqref{eq:force-cov} to assure the correct equilibrium behavior, see App.~\ref{app:relaxation}.
For a sphere of radius $R$, it is given by
\begin{align}
\label{eq:activeTractionForce}
\mathbf f_{\rm s}
=(\beta-2)\frac{K}{R^2}\left[\mathbf Q:(\bm D\mathbf Q)-\mathbf Q\cdot(\bm D\cdot\mathbf Q)\right].
\end{align}
The contraction in the first term is $[\mathbf Q:(\bm D\mathbf Q)]_j=Q^{ik}D_iQ_{kj}$.
% \begin{align}
% \label{eq:activeTractionForce}
% \mathbf f_{\rm s} &=\beta[\mathbf Q:(\bm D\mathbf Q)-\mathbf Q\cdot(\bm D\cdot\mathbf Q)]
% \end{align}
% with amplitude $\beta=-2K/R^2$. 
The same tensorial structure appears among the nonlinear active forces used to explain cell accumulation around integer defects on patterned flat substrates~\cite{Zhao2025}.
%Incidentally, this expression was introduced in Ref.~\cite{Zhao2025} to account for the orientational order of neural progenitor cells observed on microfabricated patterned flat surfaces.
We use this term without constraints on the value of $\beta$ and consider it to include an active traction force if $\beta\neq0$. %-2K/R^2$. 
For a cell, such a force would be transmitted to the surface through adhesion molecules.

We consider two limiting mechanical cases. For an incompressible fluid, $\bm D\cdot\mathbf v=0$ and $P$ acts as a Lagrange multiplier. In the perfectly compressible limit, we omit density-dependent pressure, and the isotropic stress contains elastic, viscous, and active contributions: 
\begin{align}
-P &= f - \frac{K}{2}(\bm D\mathbf Q)^2 + \eta_{\rm b}\bm D\cdot\mathbf{v} + \alpha_i S^2,
\end{align}
where $\eta_{\rm b}$ is the bulk viscosity and $\alpha_i$ the isotropic active stress coefficient. For simplicity, we set $\eta_{\rm b}=\eta$. %The passive limit is $\alpha=\alpha_i=\beta+2K/R^2=0$.

We use spherical coordinates $\{\theta,\phi\}$ with $\theta$ the colatitude measured from the north pole.
%\blue{On Mercator plots, the north pole is at the bottom.}
All vectors and tensor components are expressed in the local orthonormal basis $\{\hat{\mathbf e}_{\theta},\hat{\mathbf e}_{\phi}\}$. Explicit expressions for the covariant derivatives are given in App.~\ref{app:spherical}. For numerical computations, we expand the dynamic fields $\mathbf{v}$ and $\mathbf{Q}$ in terms of spin-weighted spherical harmonics, App.~\ref{app:swsh}.
%We show in App.~\ref{app:relaxation} that, on a sphere, the system relaxes to thermodynamic equilibrium for the parameters $\alpha=\alpha_i=\beta+2K/R^2=0$. A non-zero force density $\mathbf f_s\neq\bm 0$ is necessary for thermodynamic consistency, because of the non-commutativity of covariant derivatives, see App.~\ref{app:relaxation}. A correction force was also obtained in Ref.~\cite{Bouck2023} with different model assumptions. 
%In the presence of activity, that is, $\alpha\neq0$, $\alpha_i\neq 0$, or $\beta\neq -2K/R^2$, the system can sustain permanent flows such that the defect organization can differ from the equilibrium case.

We nondimensionalize lengths by $R$, time by $\tau=\Gamma R^2/K$, and force densities by $f_0=K/R^3$, retaining the same symbols for dimensionless quantities. Throughout, we set the dimensionless shear viscosity $\eta=1$.
We describe the nematic texture using $S$ and the director angle $\psi$, defined by $\hat{\mathbf n}=\cos(\psi)\hat{\mathbf e}_{\theta}+\sin(\psi)\hat{\mathbf e}_{\phi}$. The corresponding tensor components are $Q_{\theta\theta}=-Q_{\phi\phi}=S\cos(2\psi)$ and $Q_{\theta\phi}=S\sin(2\psi)$.

\begin{figure}
\centering
\includegraphics[width=0.9\linewidth]{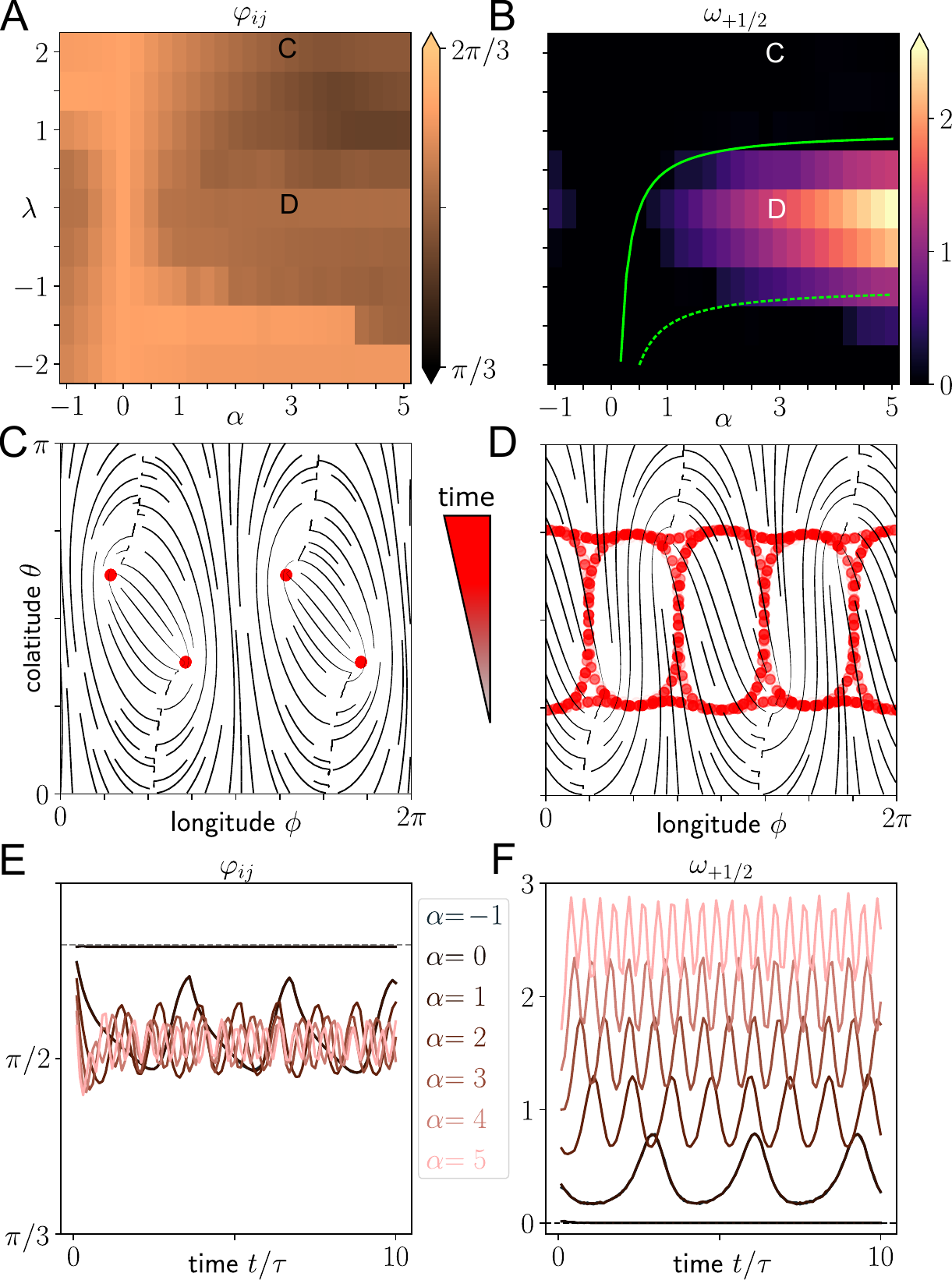}
\caption{\textit{Incompressible surface nematics with active nematic stress.}
(A,B) Characterization of the asymptotic states through the minimal defect pair angular distance $\varphi_{ij}$ (A) and the mean defect velocity $\omega_{+1/2}$ (B) as a function of the active stress and flow-alignment coefficients $\alpha$ and $\lambda$. Green lines indicate critical lines from the linear stability analysis of a flat nematic fluid, for the splay mode (full) or the bend mode (dashed).
(C,D) Mercator plots of the nematic texture (black lines) superposed with the positions of the $+1/2$-defects (red dots) for times $t=5$ (light) to $t=10$ (dark). Parameter values: $\alpha=3$ with $\lambda=2$ (C) and $\lambda=0$ (D). (E,F) Time evolution of the minimal defect-pair angle $\varphi_{ij}$ (E) and average defect angular velocity $\omega_{+1/2}$ (F) for different values of active stress $\alpha$, with $\lambda=0$. The gray dashed line in (E) corresponds to $\cos(\varphi_{ij})=-1/3$.
Other parameter values: $\chi=6$, $\xi_s=0.5$, $\beta=0$.
%\blue{invert the $\theta$ axis to get the north pole on top.}
}
\label{fig2}
\end{figure}

\paragraph*{Results.}
%In this section, we investigate the state space of incompressible or compressible surface nematics as a function of parameters.
%\subsection{Incompressible surface nematics}
We start our analysis with the incompressible case. For a passive system, $\alpha=\beta=0$, the system relaxes to thermodynamic equilibrium with four $+1/2$-defects at the vertices of a regular tetrahedron, \fig{fig1}B,C. For all pairs of defects $(i,j)$, the pairwise angular separations $\varphi_{ij}$ satisfy $\cos(\varphi_{ij})=-1/3$ or $\varphi_{ij}\simeq 1.8\pi/3$, \fig{fig2}A,E.

At $\beta=0$, active nematic stress generates spontaneous flows and propels the $+1/2$ defects, Fig.~\ref{fig2}, as in planar active nematics~\cite{Voituriez:2005jy,Giomi:2013ky}. A critical value of active stress $\alpha_{\rm c}(\lambda)$ is required to trigger spontaneous flows, which is reminiscent of the splay/bend instability where $\alpha_{\rm c}\propto -1/(\lambda\pm 1)$.
If $\lambda\lesssim 1$, we find oscillations between tetrahedral and planar defect configurations as describe previously~\cite{Keber2014}, \fig{fig2}D-F.
%The asymptotic dynamics depends strongly on the value of the flow-alignment parameter $\lambda$.
%If $|\lambda|\lesssim 1$, the system reaches oscillating states with self-propulsion of defects and spontaneous flows, \fig{fig2}B,D,F.
In contrast if $|\lambda|>1$, a static configuration appears. Strong pressure gradients balance much of the forcing, leaving only weak residual flow, \fig{figS1}.
%\textit{\blue{Comment on non-zero H}.}
For contractile activity and $\lambda>1$, we observe a pairing of two $+1/2$-defects into a vortex-like structure, \fig{fig2}A,C. Interestingly, \fig{fig2}B shows that the transition from dynamic to static states is close to the critical line from the linear stability analysis of a flat nematic fluid, $\alpha(\lambda- 1)>0$~\cite{Voituriez:2005jy}, with $\lambda_c\sim 1$.
%However, although a flat system should be unstable for $\alpha>0,\lambda<0$, on a sphere the system also transitions towards a static configuration.
However for $\alpha(\lambda\pm 1)<0$, a flat nematic fluid would undergo a spontaneous flow instability. The static configurations observed for $\lambda\lesssim-1$ indicate a reentrant transition specific to the topological properties of the sphere, \fig{fig2}A,B.

%This effect can be understood by noting first that, on a flat surface, the uniform nematic state is stable for $\alpha(\lambda- 1)>0$~\cite{Voituriez:2005jy}. The topological constraints of a sphere imply the existence of topological defects and thus intrinsic nematic gradients. For $\alpha>0$ and $\lambda>1$ these gradients result in a negative feed-back between orientation and flows, such that a static configuration is reached. Specifically, the vortex-like nematic texture balances active forces directed towards the rear of $+1/2$ comet shapes, prevents defect self-propulsion and thus leads to defect pairing/binding \blue{Not sure, which word is better}.

Nematic traction produces a different type of forcing. Consider an isolated, locally planar defect with topological charge $q$,
\begin{align}
\mathbf Q=S(\rho)
\left(2\hat{\mathbf n}\hat{\mathbf n}-\mathbf 1\right),
\end{align}
where we use planar polar coordinates $(\rho,\phi)$ and $\hat{\mathbf{n}} = \cos[(q-1)\phi + \psi_0]\hat{\mathbf{e}}_\rho + \sin[(q-1)\phi + \psi_0]\hat{\mathbf{e}}_{\phi}$. In that case
\begin{align}
\mathbf{f}_s & \propto -(\beta-2) q\,\frac{S^2(\rho)}{\rho}\,\mathbf e_\rho.
\label{eq:charge-traction}
\end{align}
The force depends on the defect charge $q$, but not on $\psi_0$, which distinguishes asters, spirals, and vortices for $q=1$ and determines the texture orientation for $q\neq1$. Thus $\beta>2$ produces inward forcing around positive defects and outward forcing around negative defects, in agreement with the case $q=1$ from~\cite{Zhao2025}.

In an incompressible fluid, pressure opposes this convergent forcing. For $\alpha=0$, the tetrahedral arrangement remains unchanged in our numerical solutions. Localized pressure maxima develop around positive defects for $\beta>2$, and minima for $\beta<2$, \fig{figS2}. The traction therefore changes the total mechanical stress without appreciably rearranging the defects.

%Let us now turn to the case of an active traction force $\mathbf f_{\rm s}$, $\beta\neq-2$. For a $+1$ defect and $\beta>0$, this force points towards the defect center independently of the nematic texture (aster, vortex, or spiral). It could thus explain the accumulation of neural progenitor cells at the core of $+1$-defects~\cite{Zhao2025}. We show in App.~\ref{app:traction-force} that accumulation occurs for any positive defect charge if $\beta>0$. Instead, depletion occurs for any negative defect charge. Note that, in the present case, incompressibility prevents any convergent flow around $+1/2$-defects. Numerical results show that in absence of active nematic stress, the active traction force does not alter the tetrahedral configuration, \fig{figS2}. A non-zero value of $\beta$ introduces frustration//mechanical stress through localized pressure monopoles around defects, with compression (dilation) for $\beta>0$ ($\beta<0$).

\begin{figure}
\centering
\includegraphics[width=0.9\linewidth]{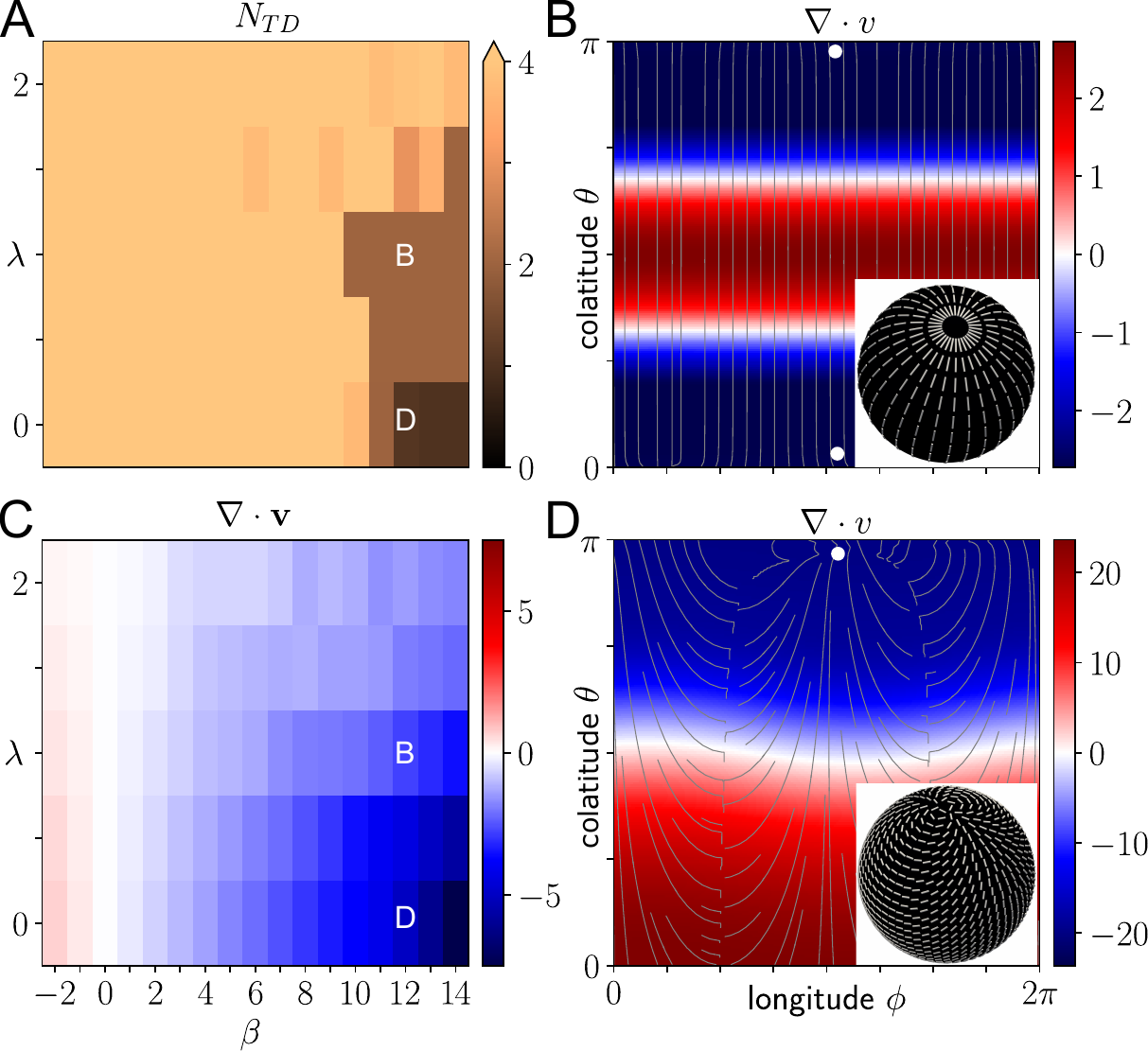}
\caption{\textit{Compressible surface nematics with active tractions.} (A,C) Phase diagrams of the defect number (A) and the flow divergence at defects (C), as a function of active traction coefficient $\beta$ and flow-alignment coefficient $\lambda$. (B,D) Mercator plots of the velocity divergence superposed with the nematic texture for $\beta=12$, $\lambda=1$ (B) and $\lambda=0$ (D). White points indicate $+1$-defects (B) or a $+2$-defect (D), and insets show three-dimensional views of the nematic texture.
Other parameters are $\chi=6$, $\alpha=\alpha_i=0$, $\xi_s=0.5$.
}
\label{fig3}
\end{figure}

%\subsection{Perfectly compressible active surface nematics}
Introducing compressibility allows a qualitatively different response. We consider the perfectly compressible limit, in which density is decoupled from force balance. At $\beta=0$, the states generated by active nematic stress remain essentially unchanged compared to the incompressible case, \fig{figS3}. For $\alpha=\alpha_i=0$, however, increasing $\beta>0$ beyond a threshold drives pairwise fusion of the four $+1/2$ defects into two $+1$ defects at opposite poles, \fig{fig3}A,B. Convergent flows entrain the defects during this process, \fig{fig3}C. In addition for $\lambda=0$, a secondary transition exists where the four $+1/2$ defects fuse into a single $+2$ defect, \fig{fig3}A,D. Beyond a threshold value $\lambda_{\rm th}\simeq 1$, the transition to $+1$-defects disappears, \fig{fig3}A, indicating that flow-alignment is important to achieve the fusion of $+1/2$-defects.
For $\beta<0$, divergent flows are obtained around defects and the topological transition is lost, \fig{fig3}A,C. It indicates that monopoles of positive flow divergence generate an effective repulsion between $+1/2$-defects.

%The flows generated by the traction force are reduced with increasing friction $\xi_s$, and defect fusion is suppressed for large enough friction [SI]. Similarly,
Increasing the nematic ordering $\chi$ localizes nematic gradients and tends to prevent defect fusion, \fig{figS4}. These observations support the idea that the convergent flows generated by the traction force lead to defect fusion.
However, substrate friction has little effect on pre-formed $+1$ defects at the poles, indicating strong metastability. In addition, low friction combined with high nematic ordering can exhibit oscillations between fused and unfused states, \fig{figS4}.
%These observations support the idea that the convergent flows generated by the traction force lead to defect fusion.

%Instead of varying the substrate friction $\xi_s$, one can vary the nematic ordering coefficient $\chi$ and modulate the spatial screening of the active force through the nematic length $\sqrt{K/\chi}$.
%In the limit where $\chi\rightarrow\infty$, nematic gradients are localized to defect cores and the velocity field vanishes in the ordered region. This prevents mechanical communication between $+1/2$-defects and cancels the transition, \fig{figS4}A. In the opposite limit $\chi\rightarrow 0$, the reorganization of the nematic texture is facilitated and the transition is promoted towards a pair of $+1$ asters, \fig{figS4}B. An nematic ordering effect also occurs through flow-alignment and convergent flows toward the poles, \fig{figS4}C,D and App.~\ref{app:traction-force}.

The mechanism of the topological transition can be understood as follows. When $+1$-defects have been formed at the poles, a flow is set along the latitude between the equator line and the defect pole, \fig{fig3}B. Assuming invariance of the nematic angle $\psi$ with respect to the azimuthal coordinate $\phi$, one writes the dominant dynamics of the nematic orientation as %in latitude coordinates 
$S\partial_t\psi\sim-\lambda\sin(2\psi)\partial_{\theta}v_{\theta}$. In first approximation, one expects the scaling $\partial_{\theta}v_{\theta}\sim(\beta-2)S^2$ by force balance, because the vanishing nematic order $S$ at the poles makes flow gradients dominant near the equator, App.~\ref{app:traction-force}. Then, the dominant orientation dynamics indicates stability of the aster configuration, $\psi=0$, for $(\beta-2)\lambda>0$, and of the vortex configuration $\psi=\pi/2$, for $(\beta-2)\lambda<0$.

\begin{figure}
\centering
\includegraphics[width=0.9\linewidth]{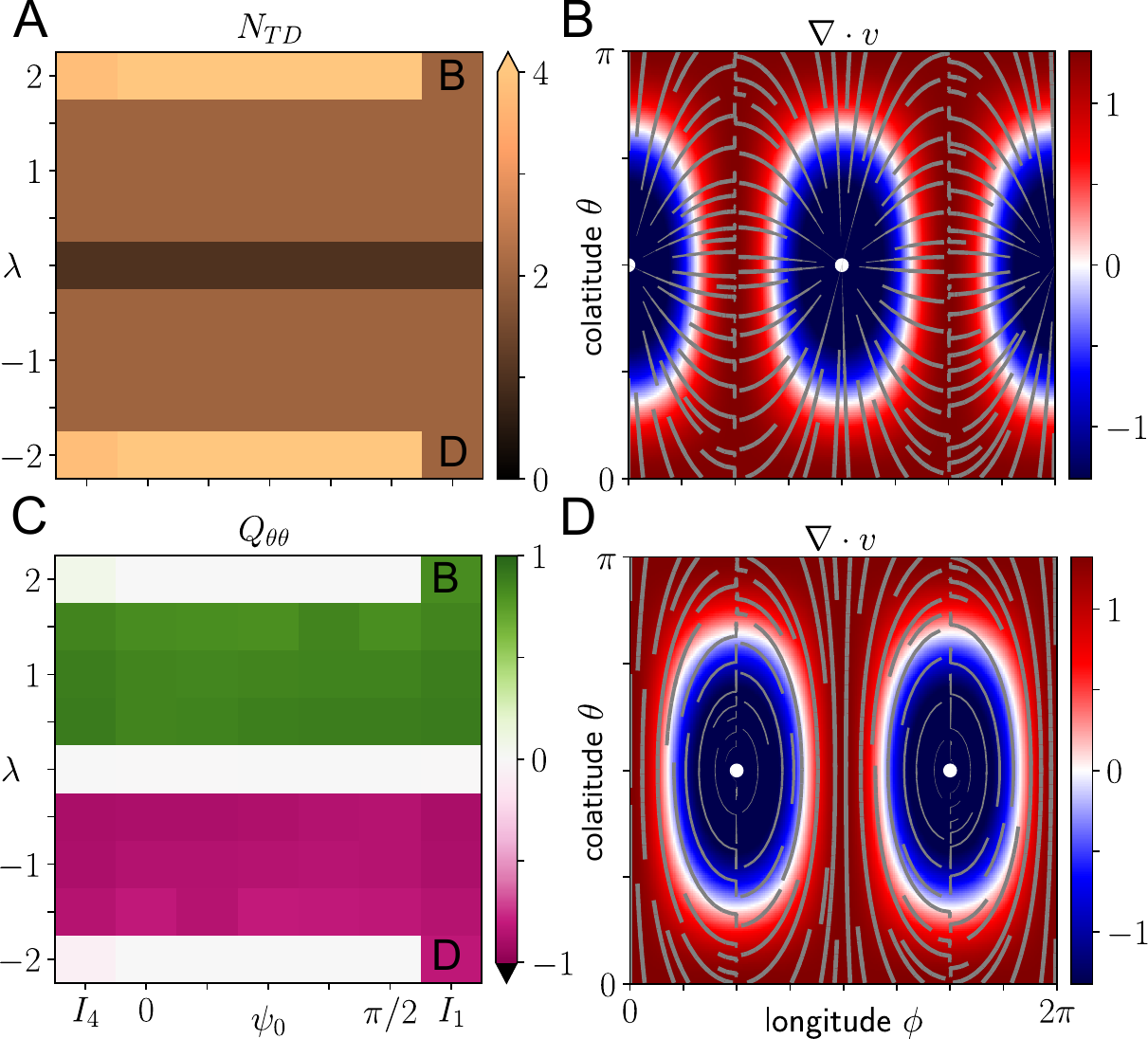}
\caption{\textit{Phase dependence of the $+1$-defect pair.}
(A,C) Phase diagrams of the defect number (A) and the mean latitude nematic $Q_{\theta\theta}$ (C) as a function of initial orientation and flow-alignment coefficient $\lambda$. Initial orientation is either the configuration $I_4$ with four $+1/2$-defects on a tetrahedron, two $+1$-defects on the pole with initial phase $\psi_0\in[0,\pi/2]$, or one $+2$-defect at the north pole $I_1$.
(B,D) Mercator plots of the velocity divergence superposed with the nematic texture at $\lambda=2$ (B) and $\lambda=-2$ (D) at $t=10$, initiated with $I_1$.
Other parameters are $\chi=6$, $\xi_s=0.5$, $\alpha=\alpha_i=0$, $\beta=14$.
}
\label{fig4}
\end{figure}

To check this argument numerically, we looked at changes of the initial orientation angle $\psi_0$ and the flow alignment coefficients $\lambda$ for $\beta=14$. We find the transition to occur for both positive and negative values of $\lambda$, \fig{fig4}A. In agreement with previous qualitative argument, one measures an average latitude nematic $Q_{\theta\theta}\simeq 1$ with aster-like orientation for positive $\lambda$, and $Q_{\theta\theta}\simeq -1$ with vortex-like orientation for negative $\lambda$, \fig{fig4}B-D. Interestingly for $\lambda=2$, quasi-bistability of a four $+1/2$-defects state and a two $+1$-defects state is observed, \fig{fig4}A.

\begin{figure}
\centering
\includegraphics[width=0.9\linewidth]{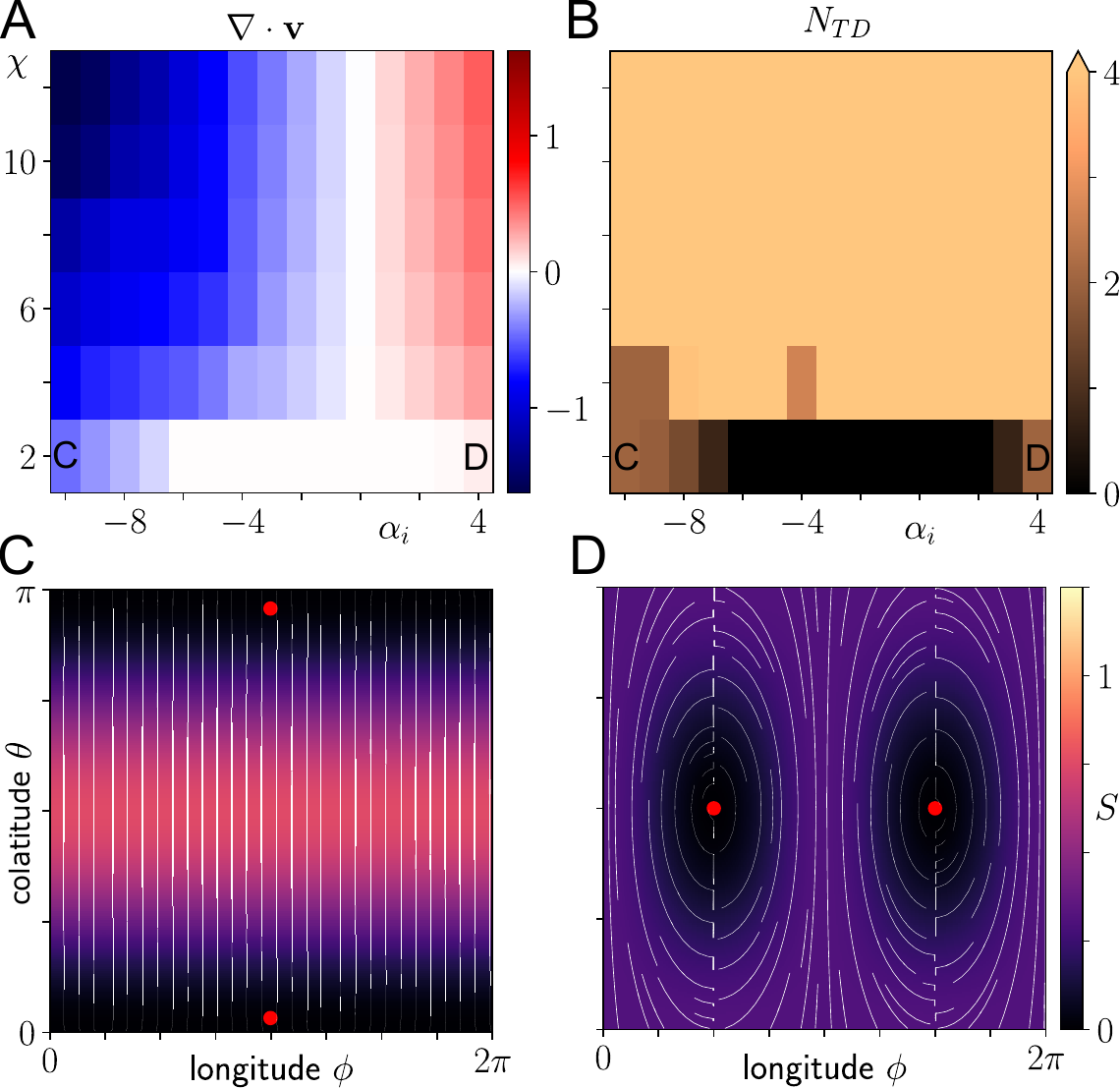}
\caption{\textit{Compressible surface nematics with active isotropic stress.}
(A,B) Phase diagrams of the velocity divergence around defects (A) and defect number (B) as a function of isotropic active stress $\alpha_i$ and nematic ordering $\chi$ coefficients. Quantities are time-averaged over $t=[5,10]$.
(C-D) Mercator plots of the nematic field (C,D) superposed with the nematic texture at $t=10$, for $\chi=2$, $\alpha_i=-10$ (C) and $\alpha_i=4$ (D). Red points indicate $+1$-defects. Other parameters are $\xi_s=0.5$, $\alpha=\beta=0$, $\lambda=1.5$.
}
\label{fig5}
\end{figure}

Finally, we looked at the effect of isotropic active stress, which generates an active force density $\mathbf f_a=2\alpha_iS\bm DS$. It also creates monopoles of velocity divergence, with inward flow for $\alpha_i<0$ since the nematic order increases from the defect core to the ordered region, \fig{fig5}A. In contrast to active tractions, the active force vanishes far from defects where gradients in nematic order are negligible. We find numerically that no topological transition exists for similar values of flow-alignment and nematic ordering, \fig{fig5}B. Only for sufficiently low values of nematic ordering $\chi$, the active force propagates to the entire system and a topological transition occurs. A similar argument based on dominant orientation dynamics shows that a contractile isotropic active stress ($\alpha_i>0$) promotes a vortex $+1$-defect, whereas the extensile case ($\alpha_i<0$) favors an aster $+1$-defect and drives nematic ordering, App.~\ref{app:traction-force}. Indeed, numerical results show an aster configuration for the extensile case with nematic ordering, \fig{fig5}C, and a vortex configuration for the contractile case, \fig{fig5}D.

\paragraph*{Discussion.}
In this work, we show that active nematic forces can spontaneously break symmetry on a sphere and establish a bipolar axis.
%In this letter, we explored a variety of configurations for a nematic fluid on a sphere driven by three active forces: a nematic stress, a nematic traction or an isotropic stress.
The nematic stress, the nematic traction, and the isotropic active stress, have all been shown to be relevant experimentally for nematic cell tissues~\cite{Zhao2025}.
By forcing a finite number of defects on a closed surface, the topology of the sphere offers a controlled system to study the role of each active force on the preferred nematic texture.
%We find that a contractile active nematic stress can induce $+1/2$-defect pairing without defect coalescence. Positive active tractions can drive a topological transition to a pair of $+1$-defect or a single $+2$ defect. Finally, the isotropic active stress can also trigger a topological transition but in a limited region of parameter space compared with active tractions.

In all cases, the orientation of the pair of $+1$-defects is sensitive to the signs of the active force and the flow-alignment coefficient. This parameter has not been measured experimentally for cell tissues, whereas it is crucial in the active nematic instability because a contractile active nematic stress $\alpha>0$ becomes unstable only for $\lambda\lesssim 0$. It also connects to a conundrum in the field, that myoblast cell monolayers behave as contractile active nematics on free substrates~\cite{Sano2017,Rembert2026} but are interpreted as extensile active nematics when confined on microchannels~\cite{Duclos2018} if $\lambda>0$. The use of a spherical topology offers an interesting experimental platform to estimate the sign of the flow-alignment coefficient $\lambda$. Our theory predicts that $\lambda>0$ ($\lambda<0$) promotes the fusion of $+1/2$-defects into a pair of aster (vortex) $+1$-defects.

The fusion of $+1/2$ defects driven by active traction and isotropic active stress requires compressibility, because incompressibility prevents local divergence in the velocity field associated with defects. As embryos are known to undergo symmetry breaking transitions during biological development, our work exemplifies the need to consider the compressibility of cell tissues. In planar systems, this compressibility might be effective when changes in the lateral extension of cells are accompanied by changes in their height such that their volume remains unchanged. It is interesting to generalize our approach and to study how thickness variations of the fluid affect the topological transitions described here.

%For the active nematic stress, we find that $+1/2$-defect self-propulsion is cancelled by pairing
% check if low chi allows defect fusion
% check if high xs cancels pairing

%\end{linenumbers}

\begin{acknowledgments}
KK gratefully acknowledges support by Swiss National Science Foundation through grant number 200020E\_219164. Numerical solutions of the dynamic equations were performed at the University of Geneva’s Baobab HPC cluster.
\end{acknowledgments}

\appendix
\setcounter{figure}{0}
\renewcommand{\thefigure}{S\arabic{figure}}

\section*{END MATTER}
\subsection*{Relaxation to equilibrium in the absence of activity}
\label{app:relaxation}

Here, we show explicitly that a nematic fluid on a sphere of radius $R$ relaxes towards thermodynamic equilibrium in the passive case $\alpha=\alpha_i=\beta=0$. To this end, we show that the total free energy $\mathcal{F}=\int\mathrm{d}A\,f$ monotonically decreases under the dynamic Eqs.~\eqref{eq:dtQ}-\eqref{eq:activeTractionForce}. Before analyzing the spherical case, we consider a fluid on a flat surface.

In flat space, the time derivative of $\mathcal{F}$ is given by
\begin{widetext}
\begin{align}
\label{eq:intermediate}
\frac{\mathrm{d}}{\mathrm{d}t}\mathcal F=-\int\mathrm{d}A\,\mathbf H:\partial_t\mathbf Q
=-\frac{1}{\Gamma}\int\mathrm{d}A\,\mathbf H^2+\int\mathrm{d}A\,\left[\mathbf H:(\mathbf v\cdot\bm\nabla)\mathbf Q+\mathbf H:(\bm\omega\cdot\mathbf Q-\mathbf Q\cdot\bm\omega)-2\lambda\mathbf H:\tilde{\bm u}\right],
\end{align}
\end{widetext}
where $\bm\nabla$ is the cartesian derivative operator. We consider the incompressible and the perfectly compressible cases separately. 

For an incompressible fluid, force balance yields $\xi_s\mathbf v-\eta\Delta\mathbf v=-\bm\nabla P+\bm\nabla\cdot(\tilde{\bm\sigma}^{\rm E}-2\lambda\mathbf H)$. Projecting this equation on $\mathbf{v}$ and integrating over space, we find $\mathcal P=\mathcal D$, where the mechanical power $\mathcal P$ and the dissipation $\mathcal D$ are given by 
\begin{align}
\mathcal P&=\int\mathrm{d}A\,\left[-\tilde{\bm\sigma}^{\rm E}+2\lambda\mathbf H\right]:\bm\nabla\mathbf v^T, \\
\mathcal D&=\int\mathrm{d}A\,\left[\xi_s\mathbf v^2+\eta\,\bm\nabla\mathbf v:\bm\nabla\mathbf v^T\right].
\end{align}
Inserting in $\mathcal P$ the expression for the traceless Ericksen stress, $\tilde{\bm\sigma}^{\rm E}=-K\bm\nabla\mathbf Q:\bm\nabla\mathbf Q+\frac{K}{2}(\bm\nabla\mathbf Q)^2\bm 1+\mathbf H\cdot\mathbf Q-\mathbf Q\cdot\mathbf H$, using the decomposition $\bm\nabla\mathbf v=\tilde{\bm u}+\bm\omega$ and performing integrations by parts, the integrand of $\mathcal P$ can be rewritten as
\begin{align}
-\mathbf H:\left(\mathbf v\cdot\bm\nabla\right)\mathbf Q-\mathbf H:(\bm\omega\cdot\mathbf Q-\mathbf Q\cdot\bm\omega)+2\lambda\mathbf H:\tilde{\mathbf u}.   
\end{align}
Inserting this result into Eq.~\eqref{eq:intermediate}, we arrive at
\begin{align}
\label{eq:dFdtincompressible}
\frac{\mathrm{d}}{\mathrm{d}t}\mathcal F=-\frac{1}{\Gamma}\int\mathrm{d}A\,\mathbf H^2\,-\mathcal D\le0.
\end{align}

For a perfectly compressible system, force balance takes the form $\bm\nabla\cdot\left(\bm\sigma^{\rm E}-2\lambda\mathbf H\right)=\xi_s\mathbf v-\eta\Delta\mathbf v-\eta_b\bm\nabla\left(\bm\nabla\cdot\mathbf v\right)$ with $\bm\sigma^{\rm E}=[f-\frac{K}{2}(\bm\nabla\mathbf Q)^2]\bm 1+\tilde{\bm\sigma}^{\rm E}$. Performing the analogous steps as in the incompressible case, we again find $\frac{\mathrm{d}}{\mathrm{d}t}\mathcal F\le0$. Compared to Eq.~\eqref{eq:dFdtincompressible}, the dissipation function $\mathcal D$ contains an additional term $+\eta_b\int\mathrm{d}A\, \left(\bm \nabla\cdot\mathbf{v}\right)^2$.

Let us now consider a nematic fluid on a sphere. We first assume incompressibility. The advection term in $d\mathcal{F}/dt$ can be written as 
\begin{align*}
I_{\rm adv}&=\int\mathrm{d}A\,\mathbf H:[(\mathbf v\cdot\bm D)\mathbf Q] \\
&=K\int\mathrm{d}A\left[D_iD_iQ_{kl}\right]v_jD_jQ_{kl} \\
&=-K\int\mathrm{d}A\left(D_iQ_{kl}\right)\left[\left(D_iv_j\right)D_jQ_{kl}+v_j[D_i,D_j]Q_{kl}\right],
\end{align*}
where $\bm D$ is the covariant derivative. For this calculation, we used incompressibility, $D_jv_j=0$. Furthermore, we introduced the commutator $[D_i,D_j]=D_iD_j-D_jD_i$ and assume summation over repeated indices. Evaluating the commutator, we get $D_iQ_{kl}[D_i,D_j]Q_{kl}=2(Q_{jk}D_iQ_{ik}-Q_{ik}D_iQ_{jk})/R^2$.  Combining the advection term with the rotational term $I_{\rm rot}=\int\mathrm{d}A\,\mathbf H:(\bm\omega\cdot\mathbf Q-\mathbf Q\cdot\bm\omega)$ and the flow-alignment term $I_{\rm flow}=-2\lambda\int\mathrm{d}A\,\mathbf H:\tilde{\mathbf u}$, we obtain
\begin{align}
I_{\rm adv}+I_{\rm rot}+I_{\rm flow}&=\int\mathrm{d}A\,[(\tilde{\bm\sigma}^{\rm E}-2\lambda\mathbf H):\bm D\mathbf v^T-\mathbf v\cdot\mathbf f_{\rm p}] \\
\mathbf f_{\rm p}&=-\frac{2K}{R^2}[\mathbf Q:\bm D\mathbf Q-\mathbf Q\cdot(\bm D\cdot\mathbf Q)].
\end{align}
Here, $\mathbf f_{\rm p}$ results from the commutator and has units of a force.

The term involving the force does not have a definite sign. To arrive at a thermodynamic consistent set of equations one could using a different expression for the stress. Equivalently, one can add an appropriate force in the covariant force balance equation. We opt for the latter and write $(\xi_s-\eta/R^2)\mathbf v-\eta D^2\mathbf v=-\bm D P+\bm D\cdot(\tilde{\bm\sigma}^{\rm E}-2\lambda\mathbf H)+\mathbf f_{\rm p}$. From this equation we obtain the associated mechanical power $\mathcal P$ and dissipation $\mathcal D$ 
\begin{align}
\mathcal P&=-\int\mathrm{d}A\,[(\tilde{\bm\sigma}^{\rm E}-2\lambda\mathbf H):\bm D\mathbf v^T-\mathbf v\cdot\mathbf f_{\rm p}] \\
\mathcal D&=\int\mathrm{d}A\,\left[(\xi_s-\tfrac{\eta}{R^2})\mathbf v^2+\eta\,\bm D\mathbf v:\bm D\mathbf v^T\right].
\end{align}
As above they satisfy $\mathcal P=\mathcal D$, such that $I_{\rm adv}+I_{\rm rot}+I_{\rm flow}=-\mathcal D$, which implies to dissipation
\begin{align}\label{eq:dFdtincompressible-cov}
\frac{\mathrm{d}}{\mathrm{d}t}\mathcal F=-\frac{1}{\Gamma}\int\mathrm{d}A\,\mathbf H^2\,-\mathcal D\le0.
\end{align}
One can check through spin-weighted harmonic expansion that $\mathcal D$ is positive-definite, see App.~\ref{app:swsh}.

For a perfectly compressible system, we write force balance as $\mathbf f_{\rm p}+\bm D\cdot\left(\bm\sigma^{\rm E}-2\lambda\mathbf H\right)=(\xi_s-\eta/R^2)\mathbf v-\eta D^2\mathbf v-\eta\bm D(\bm D\cdot\mathbf v)$. Performing the analogous steps as in the incompressible case, we again find $\frac{\mathrm{d}}{\mathrm{d}t}\mathcal F\le0$. Compared to Eq.~\eqref{eq:dFdtincompressible-cov} and similarly to the flat geometry, the dissipation function contains the additional term $+\eta_b\int\mathrm{d}A\, \left(\bm D\cdot\mathbf{v}\right)^2$.

For the active system introduced in the main text, force balance is $\bm D\cdot\bm\sigma=\xi_s\mathbf v-\mathbf f_{\rm s}$, where the total stress and the force $\mathbf f_{\rm s}$ contain active contributions. For $\alpha\neq0$, $\alpha_i\neq 0$, or $\beta\neq -2K/R^2$ the dynamics is not relaxational. Instead,  $I_{\rm adv}+I_{\rm rot}+I_{\rm flow}=-\mathcal D+\mathcal P_{\rm a}$ with
\begin{align}
\mathcal P_{\rm a}=\int\mathrm{d}A\left\{\bm D\cdot(\alpha\mathbf Q+\alpha_i\,S^2\bm 1)+\left(1+\frac{\beta R^2}{2K}\right)\mathbf f_{\rm p}\right\}\cdot\mathbf v,
\end{align}
which can have either sign.

\clearpage
\section*{SUPPLEMENTARY INFORMATION}
\section{Local defect configurations}
\label{app:traction-force}

In this Section, we explore, how the active traction force $\mathbf{f}_s$, Eq.~\eqref{eq:activeTractionForce}, affects the fields around defects. We start with analyzing a defect in a plane and then consider a defect on a sphere. We use the notation $\tilde\beta=\beta-2$.

For a defect of topological charge $q\in\mathbb{Z}/2$ in a plane, we use planar polar coordinates $(\rho,\phi)$ centered on the defect. We write the corresponding director field as $\hat{\bm n}=\cos[(q-1)\phi+\psi]\hat{\bm e}_r+\sin[(q-1)\phi+\psi]\hat{\bm e}_{\phi}$. Here, $\psi$ is a global phase, which, for $q\neq1$, determines the defect orientation. For +1 defects, $\psi=0,\pi$ corresponds to an aster, $\psi=\tfrac{1}{2}\pi, \tfrac{3}{2}\pi$ to a vortex, whereas all other values lead to spirals.

Assuming that the defect is invariant under rotations, $S\equiv S(\rho)$, the nematic tensor $\mathbf Q=S[2\hat{\bm n}\hat{\bm n}-\bm 1]$ generates a radial active traction force
\begin{equation}
\label{eq:tractionForceDefectPlane}
\mathbf f_{s}=-4\tilde\beta q\,\frac{S^2(\rho)}{\rho}\hat{\bm e}_{\rho}.
\end{equation}
This forces points towards the defect core and thus induces inward flows when $\tilde\beta q>0$. For a compressible fluid, matter correspondingly accumulates at defects with positive charge if $\tilde\beta>0$. In the opposite case, nematic tractions lead to depletion of matter.

% \section{Active traction force around defects}\label{app:traction-force}
% We consider a defect of charge $q$ and use radial coordinates $(\rho,\phi)$ centered on the defect core. The director is written as $\hat{\bm n}=\cos\theta\hat{\bm e}_x+\sin\theta\hat{\bm e}_y$ with the angle $\theta=q\phi+\psi$, where $\psi$ is a global phase. Equivalently, the director can also be written as $\hat{\bm n}=\cos[(q-1)\phi+\psi]\hat{\bm e}_r+\sin[(q-1)\phi+\psi]\hat{\bm e}_{\phi}$ in the local polar basis. Assuming a rotationally-invariant order $S(\rho)$, the nematic tensor $\mathbf Q=S[2\hat{\bm n}\hat{\bm n}-\bm 1]$ generates a radial active traction force
% \begin{equation}
% \mathbf f_s^{(\rho)}=-4\beta q S^2/r\hat{\bm e}_{\rho}
% \end{equation}
% This forces induces inward flows towards the defect core when $\beta q>0$. Then for $\beta>0$, it shows that defects of positive (negative) charge tend to exhibit material accumulation (depletion), due to active nematic tractions.

We now turn to a defect on a sphere. Without restriction of generality, we assume it to be located at the north pole, $\theta=0$, and focus on the case $q=+1$. The corresponding traction force is
\begin{align}
\mathbf f_{\rm s}=-\frac{4\tilde\beta}{R}S^2\left[\cot\theta+\frac{\partial_{\phi}\psi}{\sin\theta}\right]\hat{\mathbf e}_{\theta}+\frac{4\tilde\beta}{R}S^2\left(\partial_{\theta}\psi\right)\hat{\mathbf e}_{\phi}.
\end{align}
In the limit $\theta\to0$, this expression coincides with that in Eq.~\eqref{eq:tractionForceDefectPlane} for $q=1$, $\psi=const$, and $R\sin\theta=\rho$.

For further analysis, we consider a state with rotational invariance, $\partial_\phi=0$, and no azimuthal flows, $v_\phi=0$. Furthermore, we choose $\xi_s=0$ for simplicity, such that there is no friction between the fluid and the spherical surface. Then, force balance gives in the case of a perfectly compressible fluid
\begin{align}
\tfrac{\eta}{R^2}\left[\partial_{\theta}^2+\cot\theta\partial_{\theta}-\cot^2\theta\right]v_{\theta}= \tfrac{4\tilde\beta}{R}S^2\cot\theta.
\end{align}
%\textcolor{blue}{Are there no contributions from flow alignment etc.?}
The nematic angle and order parameter evolve according to
\begin{align}
2S\partial_t\psi& =-\frac{\lambda}{R}\sin(2\psi)(\partial_{\theta}v_{\theta}-\cot\theta v_{\theta}) \\
\partial_tS&= \frac{\lambda}{R}\cos(2\psi)(\partial_{\theta}v_{\theta}-\cot\theta v_{\theta}).
\end{align}

Close to the north pole, $\theta\in\left[0,\theta_m\right]$, we can map the sphere onto a flat disk with radial coordinate $\rho=R\theta\in\left[0,\rho_m\right]$. Under this mapping $v_\rho=v_\theta$ and
\begin{align}
\eta\left[\partial_{\rho}^2+\tfrac{1}{\rho}\partial_{\rho}-\tfrac{1}{\rho^2}\right]v_{\rho}&= \tfrac{4\tilde\beta}{\rho}S^2 \\
2S\partial_t\psi &= -\lambda\sin(2\psi)\left(\partial_{\rho}v_{\rho}-\tfrac{v_{\rho}}{\rho}\right) \\
\partial_tS&=\lambda\cos(2\psi)\left(\partial_{\rho}v_{\rho}-\tfrac{v_{\rho}}{\rho}\right)
\end{align}
%With the \textit{ansatz} $S\left(\rho\right)=\rho/\rho_m$

% For a $+1$-defect at a pole $\theta=0$ of a spherical surface, one finds similarly that
% \begin{align}
% \mathbf f_{\rm s}=-\frac{4\beta}{r}S^2\left[\cot\theta+\frac{\partial_{\phi}\psi}{\sin\theta}\right]\hat{\mathbf e}_{\theta}+\frac{4\beta}{r}S^2\partial_{\theta}\psi\hat{\mathbf e}_{\phi}
% \end{align}
% It coincides with the disc formula for $q=1$, $\psi=\psi_0$ and $\rho\rightarrow r\sin\theta$ in the limit $\theta\rightarrow 0$. Neglecting viscous friction for simplicity, local force balance around this defect is approximately $\eta D^2\mathbf v+\eta\bm D(\bm D\cdot\mathbf v)\simeq -\mathbf f_{\rm s}$. Assuming a rotationally-invariant flow $\partial_{\phi}=v_{\phi}=0$ with $v_{\theta}=0$ at $\theta=0,\pi/2$, one obtains in spherical coordinates
% \begin{align}
% \frac{\eta}{r^2}[\partial_{\theta}^2+\cot\theta\partial_{\theta}-\cot^2\theta]v_{\theta}\simeq \frac{4\beta}{r}S^2\cot\theta
% \end{align}
% Under the previous hypotheses, the shear rate is reduced to $\tilde{\mathbf u}=\frac{1}{2r}(\partial_{\theta}v_{\theta}-\cot\theta v_{\theta})(\hat{\mathbf e}_{\theta}\hat{\mathbf e}_{\theta}-\hat{\mathbf e}_{\phi}\hat{\mathbf e}_{\phi})$ and the vorticity tensor vanishes. For negligible gradients of nematic angle, one finds the orientation and order dynamics
% \begin{align}
% 2S\partial_t\psi& \simeq -\frac{\lambda}{r}\sin(2\psi)(\partial_{\theta}v_{\theta}-\cot\theta v_{\theta}) \\
% \partial_tS&\simeq \frac{\lambda}{r}\cos(2\psi)(\partial_{\theta}v_{\theta}-\cot\theta v_{\theta})
% \end{align}
For analytical purposes, we use a topological transformation from the hemisphere $\theta\in[0,\pi/2]$ to a flat disc with radial coordinate $\rho\simeq r\theta\in[0,\rho_E]$. The previous equations are mapped to
\begin{align}
\eta[\partial_{\rho}^2+\frac{\partial_{\rho}}{\rho}-\frac{1}{\rho^2}]v_{\rho}&\simeq \frac{4\tilde\beta}{\rho}S^2 \\
2S\partial_t\psi &\simeq -\lambda\sin(2\psi)\left(\partial_{\rho}v_{\rho}-\frac{v_{\rho}}{\rho}\right) \\
\partial_tS&\simeq\lambda\cos(2\psi)\left(\partial_{\rho}v_{\rho}-\frac{v_{\rho}}{\rho}\right)
\end{align}
Using the order ansatz $S(\rho)=\rho/\rho_E$, one obtains the analytical velocity profile $v_{\rho}(\rho)\simeq \frac{\tilde\beta}{2\eta\rho_E^2}\rho(\rho^2-\rho_E^2)$ with the shear rate $\tilde{u}_{\rho\rho}=\frac{1}{2}(\partial_{\rho}v_{\rho}-v_{\rho}/\rho)\simeq \frac{\tilde\beta}{2\eta}(\rho/\rho_E)^2$. This leads to orientation and order dynamics
\begin{align}
\partial_t\psi &\simeq -\sin(2\psi)\frac{\tilde\beta\lambda}{2\eta}\,\frac{\rho}{\rho_E} \\
\partial_tS &\simeq \cos(2\psi)\frac{\tilde\beta\lambda}{\eta}(\rho/\rho_E)^2
\end{align}
The first equation predicts the stability of $+1$-aster $\psi=0$ (vortex $\psi=\pi/2$) defect for $\tilde\beta\lambda>0$ ($\tilde\beta\lambda<0$). The second equation predicts a nematic ordering effect in presence of convergent flows ($\tilde\beta>0$) and $\lambda>0$.
%, which is observed on \fig{figS4}C,D.

This analysis can be repeated for the active isotropic stress under the same hypotheses, with the force density $\mathbf f_{\rm a}=2\alpha_i\,S\partial_{\theta}S/r\,\hat{\mathbf e}_{\theta}$ in spherical coordinates. In radial coordinates it becomes $\mathbf f_{\rm a}\simeq 2\alpha_i\,S\partial_{\rho}S\,\hat{\mathbf e}_{\rho}$ and one finds the velocity profile $v_{\rho}(\rho)\simeq-\frac{\alpha_i}{4\eta\rho_E^2}\rho(\rho^2-\rho_E^2)$ for the order ansatz $S(\rho)=\rho/\rho_E$. This leads to orientation and order dynamics
\begin{align}
2S\partial_t\psi &\simeq \sin(2\psi)\frac{\alpha_i\lambda}{2\eta}\,(\rho/\rho_E)^2 \\
\partial_tS&\simeq -\cos(2\psi)\frac{\alpha_i\lambda}{2\eta}(\rho/\rho_E)^2
\end{align}
Here for positive flow-alignment coefficient $\lambda$, one finds that a contractile isotropic active stress ($\alpha_i>0$) promotes a vortex $+1$-defect. Instead, an extensile isotropic active stress ($\alpha_i<0$) promotes an aster $+1$-defect and induces nematic ordering.

\section{Spherical coordinates}\label{app:spherical}

For analysis of the dynamic equations~\eqref{eq:dtQ}-\eqref{eq:activeTractionForce} on a sphere with radius $R$, we use spherical coordinates $\{\theta,\phi\}$, where $\theta=0$ at the north pole, and the corresponding orthonormal basis $\{\hat{\mathbf e}_{\theta},\hat{\mathbf e}_{\phi}\}$. In the Cartesian basis of $\mathbb{R}^3$,
\begin{align}
\hat{\mathbf e}_{\theta} &=(\cos\phi\cos\theta,\sin\phi\cos\theta,-\sin\theta)\\
\hat{\mathbf e}_{\phi} &=(-\sin\phi,\cos\phi,0).
\end{align}
with covariant connection relations $\partial_{\phi}\hat{\mathbf e}_{\phi}=-\cos\theta\hat{\mathbf e}_{\theta}$, $\partial_{\phi}\hat{\mathbf e}_{\theta}=\cos\theta\hat{\mathbf e}_{\phi}$ and $\partial_{\theta}\hat{\mathbf e}_{\phi}=\partial_{\theta}\hat{\mathbf e}_{\theta}=\bm 0$.
For convenience, we provide in the following explicit expressions for the various terms in the dynamic Eqs.~\eqref{eq:dtQ}-\eqref{eq:activeTractionForce}.
%they have components $\hat{\mathbf e}_{\theta}=(\cos\phi\cos\theta,\sin\phi\cos\theta,-\sin\theta)$ and $\hat{\mathbf e}_{\phi}=(-\sin\phi,\cos\phi,0)$. The surface normal vector is $\hat{\bm\nu}=(\cos\phi\sin\theta,\sin\phi\sin\theta,\cos\theta)$. 
% These vectors are related through
% \begin{align}
% \partial_{\phi}\hat{\mathbf e}_{\phi}&=-\sin\theta\hat{\bm\nu}-\cos\theta\hat{\mathbf e}_{\theta}\\
% \partial_{\phi}\hat{\mathbf e}_{\theta}&=\cos\theta\hat{\mathbf e}_{\phi}\\
% \partial_{\phi}\hat{\bm\nu}&=\sin\theta\hat{\mathbf e}_{\phi}\\
% \partial_{\theta}\hat{\mathbf e}_{\phi}&=\bm 0\\
% \partial_{\theta}\hat{\mathbf e}_{\theta}&=-\hat{\bm\nu}\\
% \partial_{\theta}\hat{\bm\nu}&=\hat{\mathbf e}_{\theta}.  
% \end{align}

The tangential velocity and nematic order fields can be written as 
\begin{align}
\mathbf v&=v_{\theta}\hat{\mathbf e}_{\theta}+v_{\phi}\hat{\mathbf e}_{\phi}\\
\mathbf{Q}&=Q_{\theta\theta}(\hat{\mathbf e}_{\theta}\hat{\mathbf e}_{\theta}-\hat{\mathbf e}_{\phi}\hat{\mathbf e}_{\phi})+Q_{\theta\phi}(\hat{\mathbf e}_{\theta}\hat{\mathbf e}_{\phi}+\hat{\mathbf e}_{\phi}\hat{\mathbf e}_{\theta}).
\end{align}
Here, the components of the nematic tensor can be written as a function of the nematic order parameter $S$ and the director angle $\psi$ with $Q_{\theta\theta}=-Q_{\phi\phi}=S\cos(2\psi)$ and $Q_{\theta\phi}=S\sin(2\psi)$.

We need various derivatives of the fields appearing in the dynamic equations. Let $\bm D$ denote the covariant derivative. Then 
\begin{align}
\bm DP&=\frac{1}{R}\left(\partial_{\theta}P\right)\hat{\mathbf e}_{\theta}+\frac{1}{R\sin\theta}\left(\partial_{\phi}P\right)\hat{\mathbf e}_{\phi}
\\
D^2P&=\frac{1}{R^2}\left[\partial_{\theta}^2P+\cot\theta\partial_{\theta}P+\frac{\partial_{\phi}^2P}{\sin^2\theta}\right].
\end{align}

For the gradient and the divergence of the tangent velocity field, we have
\begin{align}
\bm D\mathbf v&=\frac{1}{R}\partial_{\theta}v_{\theta}\hat{\mathbf e}_{\theta}\hat{\mathbf e}_{\theta}+\frac{1}{R}\left(\frac{\partial_{\phi}v_{\phi}}{\sin\theta}+\cot\theta v_{\theta}\right)\hat{\mathbf e}_{\phi}\hat{\mathbf e}_{\phi}\nonumber\\
&\quad\quad+\frac{\partial_{\theta}v_{\phi}}{R}\hat{\mathbf e}_{\theta}\hat{\mathbf e}_{\phi} +\frac{1}{R}\left(\frac{\partial_{\phi}v_{\theta}}{\sin\theta}-\cot\theta v_{\phi}\right)\hat{\mathbf e}_{\phi}\hat{\mathbf e}_{\theta}\\
\bm D\cdot\mathbf v&=\frac{\partial_{\theta}v_{\theta}}{R}+\frac{1}{R}\left(\frac{\partial_{\phi}v_{\phi}}{\sin\theta}+\cot\theta v_{\theta}\right).
\end{align}
Its Laplacian and the gradient of its divergence are
\begin{widetext}
\begin{align}
D^2\mathbf v&=\frac{1}{R^2}\left[\partial_{\theta}^2v_{\theta}+\cot\theta\partial_{\theta}v_{\theta}+\frac{\partial_{\phi}^2v_{\theta}}{\sin^2\theta}-\frac{2\cos\theta}{\sin^2\theta}\partial_{\phi}v_{\phi}-\cot^2\theta v_{\theta}\right]\hat{\mathbf e}_{\theta}\nonumber \\ 
&\quad\quad+\frac{1}{R^2}\left[\partial_{\theta}^2v_{\phi}+\cot\theta\partial_{\theta}v_{\phi}+\frac{\partial_{\phi}^2v_{\phi}}{\sin^2\theta}+\frac{2\cos\theta}{\sin^2\theta}\partial_{\phi}v_{\theta}-\cot^2\theta v_{\phi}\right]\hat{\mathbf e}_{\phi}\\
\bm D(\bm D\cdot\mathbf v)&=\frac{1}{R^2}\left[\partial_{\theta}^2v_{\theta}+\partial_{\theta}\left(\frac{\partial_{\phi}v_{\phi}}{\sin\theta}+\cot\theta v_{\theta}\right)\right]\hat{\mathbf e}_{\theta}
+\frac{1}{R^2\sin\theta}\left[\partial_{\phi\theta}v_{\theta}+\frac{\partial_{\phi}^2v_{\phi}}{\sin\theta}+\cot\theta\partial_{\phi}v_{\theta}\right]\hat{\mathbf e}_{\phi}.
\end{align}

Finally, writing the nematic tensor as $\mathbf Q=Q_{ij}\hat{\mathbf e}_i\hat{\mathbf e}_j$ with $i,j=\{\theta,\phi\}$ and $Q_{ij}=Q_{ji}$, one gets for its various derivatives
\begin{align}
\bm D\mathbf Q&=\frac{1}{R}\left[\left(\partial_{\theta}Q_{ij}\right)\hat{\mathbf e}_{\theta}+\frac{1}{\sin\theta}\left(\partial_{\phi}Q_{ij}\right)\hat{\mathbf e}_{\phi}\right]\hat{\mathbf e}_i\hat{\mathbf e}_j+\frac{2}{R}\cot\theta\hat{\mathbf e}_{\phi}\left[Q_{\theta\theta}(\hat{\mathbf e}_{\theta}\hat{\mathbf e}_{\phi}+\hat{\mathbf e}_{\phi}\hat{\mathbf e}_{\theta})+Q_{\theta\phi}(\hat{\mathbf e}_{\phi}\hat{\mathbf e}_{\phi}-\hat{\mathbf e}_{\theta}\hat{\mathbf e}_{\theta})\right]\\
\bm D\cdot\mathbf Q&=\frac{1}{R}\left[\partial_{\theta}Q_{\theta j}+\frac{1}{\sin\theta}\partial_{\phi}Q_{\phi j}\right]\hat{\mathbf e}_j
+\frac{2}{R}\cot\theta(Q_{\theta\phi}\hat{\mathbf e}_{\phi}+Q_{\theta\theta}\hat{\mathbf e}_{\theta})\\
(\mathbf v\cdot\bm D)\mathbf Q&=
\frac{1}{R}\left[v_{\theta}\partial_{\theta}Q_{ij}+v_{\phi}\frac{1}{\sin\theta}\partial_{\phi}Q_{ij}\right]\hat{\mathbf e}_i\hat{\mathbf e}_j
+v_{\phi}\frac{2}{R}\cot\theta \left[Q_{\theta\theta}(\hat{\mathbf e}_{\theta}\hat{\mathbf e}_{\phi}+\hat{\mathbf e}_{\phi}\hat{\mathbf e}_{\theta})+Q_{\theta\phi}(\hat{\mathbf e}_{\phi}\hat{\mathbf e}_{\phi}-\hat{\mathbf e}_{\theta}\hat{\mathbf e}_{\theta})\right]\\
\mathbf Q\cdot(\bm D\cdot\mathbf Q)&=\frac{1}{R}\left[\partial_{\theta}Q_{\theta j}+\frac{1}{\sin\theta}\partial_{\phi}Q_{\phi j}\right]Q_{jk}\hat{\mathbf e}_k
+\frac{2}{R}\cot\theta\,S^2\hat{\mathbf e}_{\theta}\\
\mathbf Q:(\bm D\mathbf Q)&=
\frac{1}{R}\left[Q_{\theta i}\partial_{\theta}Q_{ij}+Q_{\phi i}\frac{1}{\sin\theta}\partial_{\phi}Q_{ij}\right]\hat{\mathbf e}_j-\frac{2}{R}\cot\theta\,S^2\hat{\mathbf e}_{\theta}\\
D^2\mathbf Q&=\frac{1}{R^2}\left[\partial_{\theta}^2Q_{ij}+\cot\theta\partial_{\theta}Q_{ij}+\frac{\partial_{\phi}^2Q_{ij}}{\sin^2\theta}\right]\hat{\mathbf e}_i\hat{\mathbf e}_j
+\frac{4\cos\theta}{R^2\sin^2\theta}(\partial_{\phi}Q_{\theta\phi}+\cos\theta Q_{\theta\theta})(\hat{\mathbf e}_{\phi}\hat{\mathbf e}_{\phi}-\hat{\mathbf e}_{\theta}\hat{\mathbf e}_{\theta}) \\ \nonumber
&\quad\quad+\frac{4\cos\theta}{R^2\sin^2\theta}(\partial_{\phi}Q_{\theta\theta}-\cos\theta Q_{\theta\phi})(\hat{\mathbf e}_{\phi}\hat{\mathbf e}_{\theta}+\hat{\mathbf e}_{\theta}\hat{\mathbf e}_{\phi})
\end{align}
where summation over repeated indices is implied.
\end{widetext}

% The difference and sum of the last two expressions appear in the active traction force. In terms of the order parameter $S$ and the nematic angle $\psi$, they are
% \begin{align}
% \mathbf Q:(\bm D\mathbf Q)-\mathbf Q\cdot(\bm D\cdot\mathbf Q)&=\nonumber\\
% &\!\!\!\!\!\!\!\!\!\!\!\!\!\!\!\!\!\!-\frac{4}{r}S^2\left[\cot\theta+\frac{\partial_{\phi}\psi}{\sin\theta}\right]\hat{\mathbf e}_{\theta}+\frac{4}{r}S^2\partial_{\theta}\psi\hat{\mathbf e}_{\phi}\intertext{and}\nonumber\\
% \mathbf Q:(\bm D\mathbf Q)+\mathbf Q\cdot(\bm D\cdot\mathbf Q)&=2\,S\bm D S
% \end{align}
% Finally, for the co-rotational derivative we need
% \begin{align}
% \bm\omega\cdot\mathbf Q-\mathbf Q\cdot\bm\omega&=
% \omega_{ij}Q_{jk}(\hat{\mathbf e}_i\hat{\mathbf e}_k-\hat{\mathbf e}_k\hat{\mathbf e}_i)
% \end{align}
% with the expression $\bm\omega=\omega_{jk}\hat{\mathbf e}_j\hat{\mathbf e}_k+\omega_{\nu k}(\hat{\bm\nu}\hat{\mathbf e}_k-\hat{\mathbf e}_k\hat{\bm\nu})$ where $\omega_{\nu k}=0$ if the derivative operator is covariant instead of surfacic.

\section{Decomposition into spin-weighted spherical harmonics}\label{app:swsh}

To numerically solve the dynamic equations for an active nematic on a sphere, we use a spectral method. To this end, we project the fields and their dynamic equations on the set of orthonormal spin-weighted spherical harmonics (SWSH). The spin component allows us to decompose also vector and tensor fields in addition to scalars, which only require spherical harmonics. For integer spin $s$ and the usual parameters $\ell$ and $m$ of spherical harmonics, we denote the SWSH functions by $Y^s_{\ell,m}(\theta,\phi)$, where $|m|\leq \ell$ and $\ell\geq|s|$. A thorough introduction of SWSH and their application to tensor calculus on a sphere can be found in Ref.~\cite{Vasil2019}.

\subsection{Basic definitions and relations}

Spin-weighted spherical harmonics are defined as
\begin{widetext}
\begin{align}
Y^s_{\ell,m}(\theta,\phi)&=N_{\ell,m}^s\sin^{|m+s|}\left(\tfrac{\theta}{2}\right)\cos^{|m-s|}\left(\tfrac{\theta}{2}\right)P_{\ell-\ell_0}^{(|m+s|,|m-s|)}(\cos\theta)\mathrm{e}^{\mi m\phi},
\end{align}
\end{widetext}
where
\begin{align}
N_{\ell,m}^s&=(-1)^{\max(m,-s)}\sqrt{\frac{(2\ell+1)(\ell+\ell_0)!(\ell-\ell_0)!}{4\pi(\ell+\ell_1)!(\ell-\ell_1)!}}
\end{align}
with $\ell_0=\max(|m|,|s|)$ and $\ell_1=\min(|m|,|s|)$ and $P_n^{(a,b)}(z)$ denotes a Jacobi polynomial of degree $n$. Scalar spherical harmonics are obtained for $s=0$. For any spin $s$, the orthogonality relation reads
\begin{align}
\int_{0}^{2\pi}\mathrm{d}\phi\int_{0}^{\pi}\mathrm{d}\theta\,\sin\theta\,Y^s_{\ell,m}(Y^s_{\ell',m'})^*=\delta_{l,l'}\delta_{m,m'}
\end{align}
where the $*$ indicates the complex conjugate.

For the decomposition, it is convenient to change the basis in tangential space from $\{\hat{\mathbf e}_{\theta},\hat{\mathbf e}_{\phi}\}$ to $\{\hat{\mathbf e}_+,\hat{\mathbf e}_-\}$. The spin basis vectors $\hat{\mathbf e}_+$ and $\hat{\mathbf e}_-$ are defined by
\begin{equation}
\hat{\mathbf e}_+=\frac{1}{\sqrt{2}}\left(\hat{\mathbf e}_{\theta}-\mi\hat{\mathbf e}_{\phi}\right)\quad\mathrm{and}\quad
\hat{\mathbf e}_-=\frac{1}{\sqrt{2}}\left(\hat{\mathbf e}_{\theta}+\mi\hat{\mathbf e}_{\phi}\right)
\end{equation}
and satisfy
\begin{align}
(\hat{\mathbf e}_\pm)^*&=\hat{\mathbf e}_{\mp}  \\ 
\hat{\mathbf e}_+\cdot\hat{\mathbf e}_-&=\hat{\mathbf e}_-\cdot\hat{\mathbf e}_+=1\\
\hat{\mathbf e}_+\cdot\hat{\mathbf e}_+&=\hat{\mathbf e}_-\cdot\hat{\mathbf e}_-=0,
\end{align}
For a tangential vector $\mathbf v=v_\theta\hat{\mathbf e}_{\theta}+v_\phi\hat{\mathbf e}_{\phi}$, we then have $\mathbf v=V^+\hat{\mathbf e}_++V^-\hat{\mathbf e}_-$, where the spin components are related to the original components via
\begin{align}
V^{\pm}&=\frac{1}{\sqrt{2}}(v_{\theta}\pm\mi v_{\phi}).
\end{align}

Consider now a tangential tensor of rank 2, $\mathbf{A}=A^{\mu\nu}\hat{\mathbf e}_\mu\hat{\mathbf e}_\nu$, where the indices $\mu$ and $\nu$ take the values $+$ and $-$. Here and in the following, summation over repeated indices is implied unless stated otherwise. The elementary tensors in the spin basis are linked to those in the polar basis via
\begin{align}
\hat{\mathbf e}_+\hat{\mathbf e}_+&=\frac{1}{2}(\hat{\mathbf e}_{\theta}\hat{\mathbf e}_{\theta}-\hat{\mathbf e}_{\phi}\hat{\mathbf e}_{\phi})-\frac{\mi}{2}(\hat{\mathbf e}_{\theta}\hat{\mathbf e}_{\phi}+\hat{\mathbf e}_{\phi}\hat{\mathbf e}_{\theta}) \\
\hat{\mathbf e}_-\hat{\mathbf e}_-&=\frac{1}{2}(\hat{\mathbf e}_{\theta}\hat{\mathbf e}_{\theta}-\hat{\mathbf e}_{\phi}\hat{\mathbf e}_{\phi})+\frac{\mi}{2}(\hat{\mathbf e}_{\theta}\hat{\mathbf e}_{\phi}+\hat{\mathbf e}_{\phi}\hat{\mathbf e}_{\theta}) \\
\hat{\mathbf e}_-\hat{\mathbf e}_+&=\frac{1}{2}(\hat{\mathbf e}_{\theta}\hat{\mathbf e}_{\theta}+\hat{\mathbf e}_{\phi}\hat{\mathbf e}_{\phi})-\frac{\mi}{2}(\hat{\mathbf e}_{\theta}\hat{\mathbf e}_{\phi}-\hat{\mathbf e}_{\phi}\hat{\mathbf e}_{\theta}) \\
\hat{\mathbf e}_+\hat{\mathbf e}_-&=\frac{1}{2}(\hat{\mathbf e}_{\theta}\hat{\mathbf e}_{\theta}+\hat{\mathbf e}_{\phi}\hat{\mathbf e}_{\phi})+\frac{\mi}{2}(\hat{\mathbf e}_{\theta}\hat{\mathbf e}_{\phi}-\hat{\mathbf e}_{\phi}\hat{\mathbf e}_{\theta}).
\end{align}
Due to the special form of the nematic tensor $\mathbf{Q}$, we thus have 
\begin{align}
\mathbf{Q}&=Q^{++}\hat{\mathbf e}_+\hat{\mathbf e}_+ + Q^{--}\hat{\mathbf e}_-\hat{\mathbf e}_-
\end{align}
with $Q^{\pm\pm}=Q_{\theta\theta}\pm\mi Q_{\theta\phi}$.

With these relations in mind, we now give the expansions for a scalar field $P$, a tangential vector field $\mathbf{v}$, and a tangential tensor field $\mathbf{A}$ of rank 2:
\begin{align}
P(\theta,\phi)&=\sum_{\ell,m}P_{\ell,m}Y^0_{\ell,m}(\theta,\phi)\quad\mathrm{with}\quad\ell\geq 0\\
V^{\mu}(\theta,\phi)&=\sum_{\ell,m}V^{\mu}_{\ell,m}Y^{\mu}_{l,m}(\theta,\phi)\quad\mathrm{with}\quad\ell\geq1\\
A^{\mu\nu}(\theta,\phi)&=\sum_{\ell,m}A^{\mu\nu}_{\ell,m}Y^{\mu+\nu}_{\ell,m}(\theta,\phi)\quad\mathrm{with}\quad\ell\geq|\mu+\nu|,
\end{align}
where $\mu,\nu=+,-$ and no implicit summation with respect to $\mu$ or $\nu$. For evaluation of the sum $\mu+\nu$, we consider $\pm$ to be equal to $\pm1$. The summation with respect to $m$ is over all integers $m$ satisfying $|m|\le\ell$. 

The SWSH functions $Y^s_{\ell,m}$ as well as the coefficients in the above expansions are complex-valued, whereas the physical fields are real-valued. This implies $V^{-}=(V^{+})^*$ and $Q^{--}=(Q^{++})^*$. Taking into account that $Y^{-s}_{\ell,-m}=(-1)^{s+m}(Y^s_{\ell,m})^*$, the components should thus satisfy $P^{0}_{\ell,m}=(-1)^m(P^0_{\ell,-m})^*$, $V^{-s}_{\ell,-m}=(-1)^{s+m}(V^s_{\ell,m})^*$, and $A^{-\mu,-\nu}_{\ell,-m}=(-1)^{\mu+\nu+m}(A^{\mu\nu}_{\ell,m})^*$.  As a consequence, coefficients of negative spins can be fully expressed in terms of the coefficients of positive spin.

The covariant derivative operator can be written as $\bm D=\hat{\mathbf e}_{\mu}\nabla_{\nu}$. For the spin basis vectors, one has~\cite{Vasil2019}
\begin{align}
\nabla_{\mu}\hat{\mathbf e}_{\nu}=-\mu\nu\frac{\cot\theta}{\sqrt{2}R}\hat{\mathbf e}_{\nu},
\end{align}
where again $R$ is the radius of the sphere and $\mu,\nu=\{+,-\}$. In the product, $\mu$ and $\nu$ are taken to be equal to $\pm1$. Furthermore,
\begin{align}
\nabla_{\mu}Y^s_{\ell,m}&=\frac{k^{\mu}_{\ell,s}}{R}Y^{\mu+s}_{\ell,m}+\mu s\frac{\cot\theta}{\sqrt{2}R}Y^s_{\ell,m}
\end{align}
with $k^{\mu}_{\ell,s}=-\mu\sqrt{(\ell-\mu s)(\ell+\mu s+1)/2}$. This implies
\begin{align}
\nabla_\mu Y_{\ell,m}^0 &=k_{\ell,0}^\mu Y_{\ell,m}^\mu\\
\nabla_\mu\left(Y_{\ell,m}^\pm\hat{\mathbf e}_\pm\right)&=k_{\ell,\pm1}^\mu Y_{\ell,m}^{\mu\pm1}\hat{\mathbf e}_\pm\\
\nabla_\mu\left(Y_{\ell,m}^{\pm2}\hat{\mathbf e}_\pm\hat{\mathbf e}_\pm\right)&=k_{\ell,\pm2}^\mu Y_{\ell,m}^{\mu\pm2}\hat{\mathbf e}_\pm\hat{\mathbf e}_\pm\\
\nabla_\mu\left(Y_{\ell,m}^0\hat{\mathbf e}_+\hat{\mathbf e}_-\right)&=k_{\ell,0}^\mu Y_{\ell,m}^{\mu}\hat{\mathbf e}_+\hat{\mathbf e}_-.
\end{align}

\subsection{Application to an incompressible active nematic}

In the following, we present the decomposition of the dynamic equations into SWSH. We start by collecting a few useful expressions. First, we consider derivatives of the pressure $P$. We need 
\begin{align}
\bm DP&=\sqrt{\tfrac{1}{2}\ell(\ell+1)}P_{\ell,m}\left(Y^{-}_{\ell,m}\hat{\mathbf e}_--Y^{+}_{\ell,m}\hat{\mathbf e}_+\right)\\
D^2P&=-\ell(\ell+1)P_{\ell,m}Y^0_{\ell,m},
\end{align}
where from now summation is implied on repeated indices $\ell$ and $m$.

We now turn to the tangential velocity field $\mathbf{v}$. We find
\begin{align}
\label{eq:Ddotv}
\bm D\cdot\mathbf v&=\sqrt{\tfrac{1}{2}\ell(\ell+1)}\left(V^{+}_{\ell,m}-V^{-}_{\ell,m}\right)Y^0_{\ell,m}\\
D^2\mathbf v&=-(\ell^2+\ell-1)\left(V^{+}_{\ell,m}Y^{+}_{\ell,m}\hat{\mathbf e}_{+}+V^{-}_{\ell,m}Y^{-}_{\ell,m}\hat{\mathbf e}_{-}\right)\\
\bm D(\bm D\cdot\mathbf v)&=\frac{1}{2}\ell(\ell+1)\left(V^{-}_{\ell,m}-V^{+}_{\ell,m}\right)\left(Y^{+}_{\ell,m}\hat{\mathbf e}_{+}-Y^{-}_{\ell,m}\hat{\mathbf e}_{-}\right).
\end{align}
The symmetric and antisymmetric part of the strain rate tensor $\bm D\mathbf{v}$ are
\begin{align}
\tilde{\mathbf{u}}&=\sqrt{\tfrac{1}{2}(\ell-1)(\ell+2)}\left(V^{-}_{\ell,m}Y^{-2}_{\ell,m}\hat{\mathbf e}_{-}\hat{\mathbf e}_{-}-V^{+}_{\ell,m}Y^{+2}_{\ell,m}\hat{\mathbf e}_{+}\hat{\mathbf e}_{+}\right)\\
\bm\omega&=\sqrt{\tfrac{1}{8}\ell(\ell+1)}\left(V^{+}_{\ell,m}+V^{-}_{\ell,m}\right)Y^0_{\ell,m}\left(\hat{\mathbf e}_{-}\hat{\mathbf e}_{+}-\hat{\mathbf e}_{+}\hat{\mathbf e}_{-}\right)
\end{align}
The divergence of the stress tensor $\bm\sigma=S^{\mu\nu}_{\ell,m}Y^{\mu+\nu}_{\ell,m}\hat{\mathbf e}_{\mu}\hat{\mathbf e}_{\nu}$ reads
\begin{align}
\bm D\cdot\bm\sigma&=\sqrt{\tfrac{1}{2}l(l+1)}\left(S^{+-}_{\ell,m}Y^{-}_{\ell,m}\hat{\mathbf e}_{-}-S^{-+}_{\ell,m}Y^{+}_{\ell,m}\hat{\mathbf e}_{+}\right) \\ \nonumber
&+\sqrt{\tfrac{1}{2}(l-1)(l+2)}\left(S^{++}_{\ell,m}Y^{+}_{\ell,m}\hat{\mathbf e}_{+}-S^{--}_{\ell,m}Y^{-}_{\ell,m}\hat{\mathbf e}_{-}\right).
\end{align}

Finally, we give the expressions for the various derivatives of the nematic tensor $\mathbf{Q}$ in increasing tensorial order:
\begin{widetext}
\begin{align}
\bm D\cdot(\bm D\cdot\mathbf Q)&=\tfrac{1}{2}\sqrt{(\ell-1)\ell(\ell+1)(\ell+2)}\left(Q^{++}_{\ell,m}+Q^{--}_{\ell,m}\right)Y^0_{\ell,m}\\
\bm D\cdot(\bm D\cdot( D^2\mathbf Q))&=-\tfrac{1}{2}(\ell^2+\ell-4)\sqrt{(\ell-1)\ell(\ell+1)(\ell+2)}\left(Q^{++}_{\ell,m}+Q^{--}_{\ell,m}\right)Y^0_{\ell,m}\\
\bm D\cdot\mathbf Q&=\sqrt{\tfrac{1}{2}(\ell-1)(\ell+2)}\left(Q^{++}_{\ell,m}Y^{+}_{\ell,m}\hat{\mathbf e}_{+}-Q^{--}_{\ell,m}Y^{-}_{\ell,m}\hat{\mathbf e}_{-}\right)\\
D^2\mathbf Q&=-(\ell^2+\ell-4)\left(Q^{++}_{\ell,m}Y^{+2}_{\ell,m}\hat{\mathbf e}_{+}\hat{\mathbf e}_{+}+Q^{--}_{\ell,m}Y^{-2}_{\ell,m}\hat{\mathbf e}_{-}\hat{\mathbf e}_{-}\right)\\
\bm D\mathbf Q&=\sqrt{\tfrac{1}{2}(\ell-1)(\ell+2)}\left(Q^{++}_{\ell,m}Y^{+}_{\ell,m}\hat{\mathbf e}_{-}\hat{\mathbf e}_{+}\hat{\mathbf e}_{+}-Q^{--}_{\ell,m}Y^{-}_{\ell,m}\hat{\mathbf e}_{+}\hat{\mathbf e}_{-}\hat{\mathbf e}_{-}\right) \nonumber\\
&\quad\quad+\sqrt{\tfrac{1}{2}(\ell-2)(\ell+3)}\left(Q^{--}_{\ell,m}Y^{-3}_{\ell,m}\hat{\mathbf e}_{-}\hat{\mathbf e}_{-}\hat{\mathbf e}_{-}-Q^{++}_{\ell,m}Y^{+3}_{\ell,m}\hat{\mathbf e}_{+}\hat{\mathbf e}_{+}\hat{\mathbf e}_{+}\right)\\
\bm D(D^2\mathbf Q)&=-(\ell^2+\ell-4)\sqrt{\tfrac{1}{2}(\ell-1)(\ell+2)}\left(Q^{++}_{\ell,m}Y^{+}_{\ell,m}\hat{\mathbf e}_{-}\hat{\mathbf e}_{+}\hat{\mathbf e}_{+}-Q^{--}_{\ell,m}Y^{-}_{\ell,m}\hat{\mathbf e}_{+}\hat{\mathbf e}_{-}\hat{\mathbf e}_{-}\right)\nonumber \\
&\quad\quad-(\ell^2+\ell-4)\sqrt{\tfrac{1}{2}(\ell-2)(\ell+3)}\left(Q^{--}_{\ell,m}Y^{-3}_{\ell,m}\hat{\mathbf e}_{-}\hat{\mathbf e}_{-}\hat{\mathbf e}_{-}-Q^{++}_{\ell,m}Y^{+3}_{\ell,m}\hat{\mathbf e}_{+}\hat{\mathbf e}_{+}\hat{\mathbf e}_{+}\right)
\end{align}

Using these results, we give the spectral form of the dynamic equations. The tangential force balance reads
\begin{align}\label{eq:force-cov1}
-\bm DP+(\alpha-2\lambda\chi)\bm D\cdot\mathbf Q-2\lambda\bm D\cdot(D^2\mathbf Q)+\mathbf f_{\rm NL}&=(\xi_{\rm s}-\eta)\mathbf v-\eta D^2\mathbf v,
\end{align}
where $\mathbf f_{NL}$ contains all nonlinear terms in $\mathbf{Q}$. We will discuss this term in detail below. Suffice it for now to remark that its decomposition into SWSH is similar to that of the velocity field, $\mathbf f_{NL}=f^{+}\hat{\mathbf e}_{+}+f^{-}\hat{\mathbf e}_{-}$ with $f^{\pm}=f^{\pm}_{\ell,m}Y^{\pm}_{\ell,m}$. By applying the divergence operator to Eq.~\eqref{eq:force-cov1} and using the incompressibility condition $\bm D\cdot\mathbf{v}=0$, we obtain
\begin{align}
\label{eq:D2P}
D^2P&=(\alpha-2\lambda\chi)\bm D\cdot(\bm D\cdot\mathbf Q)-2\lambda\bm D\cdot\bm D\cdot D^2\mathbf Q+\bm D\cdot\mathbf f_{NL}.
\end{align}
% \intertext{where}
% \mathbf f_{NL}&=\bm D\cdot\tilde{\bm\sigma}^{\rm E}+2\lambda\chi\,S^2\bm D\cdot\mathbf Q+4\lambda\chi\,S\bm DS\cdot\mathbf Q+\beta[\mathbf Q:(\mathbf D\mathbf Q)-\mathbf Q\cdot(\mathbf D\cdot\mathbf Q)]
% \end{align}
\end{widetext}
We can solve Eqs.~\eqref{eq:force-cov1},\eqref{eq:D2P} explicitly and get
\begin{widetext}
\begin{align}
[\xi_s+\eta(\ell^2+\ell-2)]V^{+}_{\ell,m}&=\sqrt{\tfrac{1}{2}\ell(\ell+1)}P_{\ell,m}+\left[\alpha+2\lambda(\ell^2+\ell-4-\chi)\right]\sqrt{\tfrac{1}{2}(\ell-1)(\ell+2)}Q^{++}_{\ell,m}+f^{+}_{\ell,m} \\
P_{\ell,m}&=-\tfrac{1}{2}\left[\alpha+2\lambda(\ell^2+\ell-4-\chi)\right]\sqrt{\tfrac{(\ell-1)(\ell+2)}{\ell(\ell+1)}}(Q^{++}_{\ell,m}+Q^{--}_{\ell,m})-\tfrac{1}{\sqrt{2\ell(\ell+1)}}\left(f^{+}_{\ell,m}-f^{-}_{\ell,m}\right).
\end{align}
\end{widetext}
Note that these equations hold for $\ell>0$. The coefficient $P_{0,0}$ remains unspecified, but is irrelevant as it corresponds to an additive constant in the pressure $P$.

When employing the result for $P_{\ell,m}$ in the force balance equation, we obtain after some algebra
%\begin{widetext}
\begin{align}
\label{eq:Vellmplus}
% \left[\eta(\ell^2+\ell-1)+\xi_s\right]V^{+}_{\ell,m}
% =\tfrac{1}{2}\left[\alpha+2\lambda(\ell^2+\ell-4-\chi)\right]\sqrt{\tfrac{1}{2}(\ell-1)(\ell+2)}\left(Q^{++}_{\ell,m}-Q^{--}_{\ell,m}\right)+\tfrac{1}{2}\left(f^{-}_{\ell,m}+f^{+}_{\ell,m}\right),
V^{+}_{\ell,m}
&=A_\ell\left(Q^{++}_{\ell,m}-Q^{--}_{\ell,m}\right)+B_\ell\left(f^{+}_{\ell,m}+f^{-}_{\ell,m}\right), \\
A_\ell&=B_\ell[\alpha+2\lambda(\ell^2+\ell-4-\chi)]\sqrt{\tfrac{1}{2}(\ell-1)(\ell+2)} \\
B_\ell&=\frac{1}{2}\left(\eta(\ell^2+\ell-2)+\xi_s\right)^{-1}
\end{align}
%\end{widetext}
for $\ell> 0$, and no summation over $\ell$ is intended. Furthermore, $V_{0,0}^+=0$. We have now explicitly determined all coefficients $V_{\ell,m}^+$. The values of $V_{\ell,m}^-$  follow directly from incompressibility, which imposes $V_{\ell,m}^-=V_{\ell,m}^+$, Eq.~\eqref{eq:Ddotv}. Alternatively, one might use $V^{-}_{\ell,-m}=(-1)^{m+1}(V^+_{\ell,m})^*$. It yields the same result as can be checked by inserting this relation in Eq.~\eqref{eq:Vellmplus} and using  $Q_{\ell,m}^{--}=(-1)^m\left(Q_{\ell,-m}^{++}\right)^*$. 

With $P_{\ell,m}$ and $V_{\ell,m}^\pm$ expressed through the coefficients of the tangential nematic field $\mathbf{Q}$, Eq.~\eqref{eq:dtQ} can be used to determine their time evolution. We only need the time evolution of the components $Q^{++}_{\ell,m}$, which gives after projection on the spin direction $\hat{\mathbf{e}}_+$
\begin{widetext}
\begin{align}
\label{eq:dQpplmdt}
\left[\partial_tQ^{++}_{\ell,m}+(\ell^2+\ell-4-\chi)Q^{++}_{\ell,m}\right]Y^{+2}_{\ell,m}
&=-2\lambda u^{++}+2\omega^{-+}Q^{++}+V^{-}\nabla Q^{+}-V^{+}\nabla Q^{-}-\chi|Q^{++}|^2Q^{++},
\end{align}
\end{widetext}
where we have introduced 
\begin{align}
\nabla Q^{-}&=\sqrt{\tfrac{1}{2}{(\ell-1)(\ell+2)}}Q^{++}_{\ell,m}Y^{+}_{\ell,m}\\
\nabla Q^{+}&=\sqrt{\tfrac{1}{2}{(\ell-2)(\ell+3)}}Q^{++}_{\ell,m}Y^{+3}_{\ell,m}.
\end{align}
The relevant components of the velocity gradient tensor are 
\begin{align}
u^{++}&=\sqrt{\tfrac{1}{2}{(\ell-1)(\ell+2)}}V^{+}_{\ell,m}Y^{+2}_{\ell,m}\\
\omega^{-+}&=\sqrt{\tfrac{1}{8}\ell(\ell+1)}(V^{+}_{\ell,m}+V^{-}_{\ell,m})Y^{0}_{\ell,m}.
\end{align}
To obtain the dynamic equations for the coefficients $Q^{++}_{\ell,m}$, one needs to project Eq.~\eqref{eq:dQpplmdt} on $Y_{\ell,m}^{+2}$.
%As noted above, the coefficients $Q^{--}_{\ell,m}$ can be obtained from the coefficients $Q^{++}_{\ell,m}$.

Finally, we specify $\mathbf{f}_{NL}$. In covariant form it reads
\begin{widetext}
\begin{align}
\mathbf{f}_{NL}&=\bm D\cdot\bm\sigma^{\rm NL}+\beta\left[\mathbf Q:(\mathbf D\mathbf Q)-\mathbf Q\cdot(\mathbf D\cdot\mathbf Q)\right],
\\
\bm\sigma^{\rm NL}&=-\bm D\mathbf Q:\bm D\mathbf Q+\frac{1}{2}(\bm D\mathbf Q)^2\bm 1+\mathbf H\cdot\mathbf Q-\mathbf Q\cdot\mathbf H+2\lambda\chi S^2\mathbf Q
\end{align}
\end{widetext}
One uses the spin decomposition of the molecular field $\mathbf H=H^{++}\hat{\mathbf e}_{+}\hat{\mathbf e}_{+}+H^{--}\hat{\mathbf e}_{-}\hat{\mathbf e}_{-}$ and the non-linear stress $\bm\sigma^{\rm NL}=\sigma^{\mu\nu}\hat{\mathbf e}_{\mu}\hat{\mathbf e}_{\nu}$ with the relevant components
%\begin{widetext}
\begin{align}
\sigma^{++}&=2\nabla Q^{+}(\nabla Q^{-})^*+2\lambda\chi |Q^{++}|^2Q^{++} \\
\sigma^{-+}&=(H^{++})^*Q^{++}-H^{++}(Q^{++})^* \\
H^{++}&=-(\ell^2+\ell-4-\chi)Q^{++}_{\ell,m}Y^{+2}_{\ell,m}-\chi|Q^{++}|^2Q^{++},
\end{align}
%\end{widetext}
%Here,
%\begin{align}
%\nabla S&=\sqrt{\frac{\ell(\ell+1)}{2}}S_{\ell,m}Y^{+}_{\ell,m}\\
%\Delta Q^{+}&=(\ell^2+\ell-4)Q^{++}_{\ell,m}Y^{+2}_{\ell,m}\\
%\nabla^3Q^{-}&=(\ell^2+\ell-4)\sqrt{\frac{(\ell-1)(\ell+2)}{2}}Q^{++}_{\ell,m}Y^{+}_{\ell,m}\\
%\nabla^3Q^{+}&=(\ell^2+\ell-4)\sqrt{\frac{(\ell-2)(\ell+3)}{2}}Q^{++}_{\ell,m}Y^{+3}_{\ell,m}.
%\end{align}
One obtains the divergent force $\bm D\cdot\bm\sigma^{\rm NL}$ by spectral transformation. The relevant components of the stress divergence are 
\begin{align}
\nabla\sigma^{-}&=\sqrt{\tfrac{1}{2}\ell(\ell+1)}\sigma^{-+}_{\ell,m}Y^{+}_{\ell,m}\\
\nabla\sigma^{+}&=\sqrt{\tfrac{1}{2}(\ell-1)(\ell+2)}\sigma^{++}_{\ell,m}Y^{+}_{\ell,m}.
\end{align}
Projection on $\hat{\mathbf{e}}_+$ yields the total non-linear force component
\begin{align}
f^+ &=\nabla\sigma^{+}-\nabla\sigma^{-}-\beta\left[Q^{++}(\nabla Q^{-})^*+(Q^{++})^*\nabla Q^{+}\right].
\end{align}
Projection of this equation on $Y^+_{\ell,m}$ gives the equations for the components $f^+_{\ell,m}$. As shown above, they also determine the components $f^-_{\ell,m}$ by exploiting that the active traction force has to be real.

\subsection{Numerical integration}

To solve the dynamic Eq.~\eqref{eq:dQpplmdt} for the time evolution of $Q_{\ell,m}^{++}$, we use an implicit-explicit scheme, where the linear terms are solve implicitly and the nonlinear terms explicitly. This avoids numerical instabilities that are introduced by the flow-alignment term.

Using the solution for $V^+_{\ell,m}$, Eq.~\eqref{eq:Vellmplus}, we can decompose the component of the symmetric part of the velocity gradient tensor into a part that is linear and one that is nonlinear in $\mathbf{Q}$, $u^{++}=u^{++}_\textrm{L}+u^{++}_\textrm{NL}$, where 
\begin{align}
u^{++}_\textrm{L}&=\sqrt{\tfrac12\left(\ell-1\right)\left(\ell+2\right)}A_\ell\left(Q_{\ell,m}^{++}-Q_{\ell,m}^{--}\right)Y_{\ell,m}^{+2}\\
u^{++}_\textrm{NL}&=\sqrt{\tfrac12\left(\ell-1\right)\left(\ell+2\right)}B_\ell\left(f_{\ell,m}^++f_{\ell,m}^-\right)Y_{\ell,m}^{+2}. 
\end{align}
We then rewrite Eq.~\eqref{eq:dQpplmdt} as
\begin{align}
\label{eq:dQpplmdt2}
\partial_tQ^{++}_{\ell,m}+(\ell^2+\ell-4-\chi+C_\ell)Q^{++}_{\ell,m}-C_\ell Q^{--}_{\ell,m}&=R^{++}_{\ell,m},
\end{align}
where $C_\ell=\frac{1}{2}\frac{\lambda(\ell-1)(\ell+2)}{\eta(\ell^2+\ell-2)+\xi_s}\left[\alpha+2\lambda(\ell^2+\ell-4-\chi)\right]$ and the $R_{\ell,m}^{++}$ contain all nonlinear terms in $\mathbf{Q}$. Explicitly, 
\begin{align}
R^{++}&=-2\lambda u^{++}_\textrm{NL}+2\omega^{-+}Q^{++}-\chi|Q^{++}|^2Q^{++}\nonumber\\
&\qquad+\nabla Q^{+}V^{-}-\nabla Q^{-}V^{+}\\
&\equiv R^{++}_{\ell,m}Y^{+2}_{\ell,m}.
\end{align}

At time $t$, we compute $R^{++}$ in real space. Then we obtain the coefficients $R_{\ell,m}^{++}$ by applying the SWSH transform. For the left hand side of Eq.~\eqref{eq:dQpplmdt2}, we first express $Q_{\ell,m}^{--}$ in terms of $Q_{\ell,m}^{++}$ through the projection $\hat{\bm P}[Q^{++}_{\ell,m}]=Q^{--}_{\ell,m}=(-1)^m(Q^{++}_{\ell,-m})^*$. The terms other than the time derivative are evaluated at time $t+1$. We thus arrive at
\begin{align}
\left\{a_\ell\bm 1 - b_\ell\hat{\bm P}\right\}Q_{\ell,m}^{++}(t+1) &= Q_{\ell,m}^{++}(t) + \Delta t R^{++}_{\ell,m}(t) \\ \nonumber
&\equiv B^{++}_{l,m}(t)
\end{align}
with $a_\ell=1+\Delta t(\ell^2+\ell-4-\chi+C_\ell)$ and $b_\ell=\Delta t C_\ell$. Using the identity $(a_\ell\bm 1-b_\ell\hat{\bm P})^{-1}=(a_\ell\bm 1+b_\ell\hat{\bm P})/\left(a_\ell^2-b_\ell^2\right)$ with $\hat{\bm P}^2=\bm I$, we finally obtain
\begin{align}
\label{eq:dtQ-update}
Q^{++}_{\ell,m}(t+1)&=\frac{a_\ell\,B^{++}_{\ell,m}(t)+b_\ell\,B^{--}_{\ell,m}(t)}{a_\ell^2-b_\ell^2}.
\end{align}
The eigenvalues associated to this discrete dynamics are $(a_l\pm b_l)^{-1}$ with $a_l\pm b_l=\{1+\Delta t(\ell^2+\ell-4-\chi),1+\Delta t(\ell^2+\ell-4-\chi+2C_\ell)\}$. For $\ell\geq 2$, one finds an instability $|a_l\pm b_l|^{-1}>1$ when $\chi>2$ corresponding to an isotropic-to-nematic transition on the sphere.

The numerical code is written in \textit{Python}, and the forward and backward SWSH transforms use the package \textit{pyspherical}~\cite{Lanman2020_pyspherical} for fast SWSH transforms~\cite{Trapani2006,McEwen2011}.

\subsection{Application to a perfectly compressible active nematic}

For a perfectly compressible active nematic on a sphere, the total stress tensor is
\begin{align}
\bm\sigma&=\eta(\bm D\mathbf v+\bm D\mathbf v^T)+\bm\sigma^E-2\lambda\mathbf H+\alpha\mathbf Q+\alpha_i\mathbf Q\cdot\mathbf Q,
\end{align}
where the Ericksen stress is $\bm\sigma^{\rm E}=f\bm 1-\bm D\mathbf Q:\bm D\mathbf Q+\mathbf H\cdot\mathbf Q-\mathbf Q\cdot\mathbf H$. Force balance then yields
\begin{widetext}
\begin{align}
\bm D\cdot\left(\bm\sigma^{\rm E}-2\lambda\mathbf H\right)+\alpha\bm D\cdot\mathbf Q+\left(\alpha_i+\beta\right)\mathbf Q:\left(\bm D\mathbf Q\right)+\left(\alpha_i-\beta\right)\mathbf Q\cdot\left(\bm D\cdot\mathbf Q\right)&=(\xi_s-\eta)\mathbf v-\eta D^2\mathbf v-\eta\bm D\left(\bm D\cdot\mathbf v\right).
\end{align}
For the corresponding spectral components we obtain
\begin{align}
\left[\tfrac{1}{2}\eta(3\ell^2+3\ell-4)+\xi_s\right]V^{+}_{\ell,m}-\frac{\eta}{2}\ell(\ell+1)V^{-}_{\ell,m}
&=[\alpha+2\lambda(\ell^2+\ell-4-\chi)]\sqrt{\tfrac{1}{2}(\ell-1)(\ell+2)}Q^{++}_{\ell,m}+f^{+}_{\ell,m},
%\\
%\left[\tfrac{1}{2}\eta(3\ell^2+3\ell-4)+\xi_s\right]V^{-}_{\ell,m}-\frac{\eta}{2}\ell(\ell+1)V^{+}_{\ell,m}
%&=-[\alpha+2\lambda(\ell^2+\ell-4-\chi)]\sqrt{\tfrac{1}{2}(\ell-1)(\ell+2)}Q^{--}_{\ell,m}+f^{-}_{\ell,m},
\end{align}
\end{widetext}
where we have used $f^+$ to indicate as above all terms nonlinear in $\mathbf{Q}$. The non-linear stress $\bm\sigma^{\rm NL}$ has the relevant components
\begin{align}
\sigma^{++}&=2\nabla Q^{+}(\nabla Q^{-})^*+2\lambda\chi |Q^{++}|^2Q^{++} \\
\sigma^{-+}&=-\chi S^2+\tfrac{1}{2}\chi S^4+(H^{++})^*Q^{++}-H^{++}(Q^{++})^*
\end{align}
and the non-linear force has the explicit expression
\begin{align}
f^+ &=\nabla\sigma^{+}-\nabla\sigma^{-} \\ \nonumber
&\quad+(\alpha_i-\beta)Q^{++}(\nabla Q^{-})^*-(\alpha_i+\beta)(Q^{++})^*\nabla Q^{+}.
\end{align}

Combining the force balance equation with its conjugate, we can solve for the velocity coefficients. We get
\begin{align}
V^{+}_{\ell,m}=A^{+}_\ell Q^{++}_{\ell,m}-A^{-}_\ell Q^{--}_{\ell,m}+B^{+}_\ell f^{+}_{\ell,m}+B^{-}_{\ell}f^{-}_{\ell,m}, 
\end{align}
where the coefficients are $A_\ell^\pm=D_\ell B_\ell^\pm$,
\begin{align}
B^+_\ell &= \frac{2\xi_s+\eta(3\ell^2+3\ell-4)}{2\left[\xi_s+\eta(\ell^2+\ell-2)\right]\left[\xi_s+2\eta(\ell^2+\ell-1)\right]} \\
B^-_\ell &= \frac{\eta\ell(\ell+1)}{2\left[\xi_s+\eta(\ell^2+\ell-2)\right]\left[\xi_s+2\eta(\ell^2+\ell-1)\right]}\\
D_\ell &= \left[\alpha+2\lambda\left(\ell^2+\ell-4-\chi\right)\right]\sqrt{\tfrac{1}{2}(\ell-1)(\ell+2)}.
\end{align}
Similarly to the case of an incompressible fluid, this allows us to split the symmetric part of the velocity gradient tensor into a part linear and one nonlinear in $\mathbf{Q}$. This leads to the analog of Eq.~\ref{eq:dQpplmdt2} for the perfectly compressible case:
\begin{align}
\label{eq:dQpplmdt3}
\partial_tQ^{++}_{\ell,m}+(\ell^2+\ell-4-\chi+C^{+}_\ell)Q^{++}_{\ell,m}-C^{-}_\ell Q^{--}_{\ell,m}
= R^{++}_{\ell,m},
\end{align}
where $C^{\pm}_\ell=\lambda\sqrt{2(\ell-1)(\ell+2)}\,A^{\pm}_\ell$. The quantities on the right hand side denote the expansion coefficients of the positive spin components nonlinear in $\mathbf{Q}$,
\begin{align}
R^{++} &=-2\lambda u^{++}_{\rm NL}+2\omega^{-+}Q^{++}\nonumber\\
&\qquad-\chi|Q^{++}|^2Q^{++}+\nabla Q^{+}V^{-}-\nabla Q^{-}V^{+}.
\end{align}
From Eq.~\eqref{eq:dQpplmdt3}, we obtain, as in the incompressible case, the update rule \eqref{eq:dtQ-update}, but now with $a_\ell=1+\Delta t(\ell^2+\ell-4-\chi+C^{+}_\ell)$ and $b_\ell=\Delta tC^{-}_\ell$.

\section{Data analysis}
The initial nematic configuration is either four $+1/2$-defects on a regular tetrahedron ($I_4$), two $+1$-defects at the poles or a quasi $+2$-defect at the north pole ($I_1$).
\begin{itemize}
    \item For two $+1$-defects, one uses the decomposition $\mathbf Q=S\cos(2\psi)(\hat{\mathbf e}_{\theta}\hat{\mathbf e}_{\theta}-\hat{\mathbf e}_{\phi}\hat{\mathbf e}_{\phi})+S\sin(2\psi)(\hat{\mathbf e}_{\theta}\hat{\mathbf e}_{\phi}+\hat{\mathbf e}_{\phi}\hat{\mathbf e}_{\theta})$ with the order field $S(\theta)=0.25\sin(\theta)$ and the orientation field $\psi(\theta,\phi)=\psi_0+0.1\sin(\theta)\cos(\phi)$, where $\psi_0$ parametrizes the global phase of the defect pair from aster ($\psi_0=0$) to vortex ($\psi_0=\pi/2$).
    \item For the tetrahedral configuration $I_4$, one uses a stereographic projection. A regular tetrahedron inscribed in a unit sphere has four vertices with Cartesian coordinates $\{\mathbf r_i\}=\{(\pm x_0,\pm x_0,\pm x_0)\}$ where $x_0=1/\sqrt{3}$. The coordinates of the stereographic plane $Z=X+\mathrm{i}Y$ are linked to spherical coordinates by $Z=\tan(\theta/2)\mathrm{e}^{\mathrm{i}\phi}$. The stereographic $q$-tensor is defined as $q(\theta,\phi)=\Pi_{i=1}^{4}\,f(|Z-Z_i|,|Z|)(Z-Z_i)$ where $Z_j=(x_j+\mathrm{i}y_j)/(1-z_j)$ are the stereographic coordinates of the tetrahedron vertices, and $f(d,a)=[d^2+r_c^2(1+a^2)(1+d^2)]^{-1/2}$ is a scalar function used to weight the order parameter symmetrically on the four defects ($r_c=0.2$). The spherical nematic tensor $Q^{++}=Q_{\theta\theta}+\mathrm{i}Q_{\theta\phi}$ is obtained by projection of the stereographic tensor, $Q^{++}=q\mathrm{e}^{-2\mathrm{i}\phi}$.
    \item For the $+2$-defect configuration $I_1$, one uses two nearby $+1$-defects at the north pole with cartesian coordinates $\{\mathbf r_i\}=\{(-\sin\varphi,0,-\cos\varphi),(\sin\varphi,0,-\cos\varphi)\}$, where $\varphi=0.5$ parametrizes the (half) angular distance between defects. The stereographic $q$-tensor is now defined as $q(\theta,\phi)=\Pi_{i=1}^{2}\,\tanh(k|Z-Z_i|)(Z-Z_i)/|Z-Z_i|$ where $k=5$ controls the spatial localization of nematic disorder, and the spherical nematic tensor is obtained as $Q^{++}=q\mathrm{e}^{-2\mathrm{i}\phi}$.
\end{itemize}

The detection of defects is based on order threshold. The field $S(\theta,\phi)$ is binarized from the threshold $S_{\rm th}=0.25$, and the \textit{label} function from the Python package \textit{Scipy.ndimage} is used to detect disordered regions. Each set of cluster coordinates $(\theta_{\rm c},\phi_{\rm c})$ is averaged to define the associated defect position $(\theta_{\rm d},\phi_{\rm d})$. One averages over $(\cos\phi_{\rm c},\sin\phi_{\rm c})$ to account for periodicity and obtains the mean azimuthal angle $\langle\phi_{\rm c}\rangle=\arctan(\langle\sin\phi_{\rm c}\rangle/\langle\cos\phi_{\rm c}\rangle)$.

Distance between defects is done by defining the three-dimensional radial vectors $\mathbf r_i=\sin\theta_i(\cos\phi_i\hat{\mathbf e}_x+\sin\phi_i\hat{\mathbf e}_y)+\cos\theta_i\hat{\mathbf e}_z$ of any defect $i$ with spherical coordinates $(\theta_i,\phi_i)$, and computing the scalar product $\mathbf r_i\cdot\mathbf r_j=\cos\varphi_{ij}$ for distinct defects $i,j$. The angular velocity of $+1/2$-defects is defined between two frames at times $\{t,t+\Delta t\}$ as $\omega_{+1/2}=\mathrm{Arg}[\mathbf r_t\cdot\mathbf r_{t+\Delta t}]/\Delta t$ for a given defect. Linkage of defects between frames is based on a minimal distance to the previous frame position.

For integer defects, one wants to rotate spherical coordinates so that $\hat{\mathbf e}_z'$ coincides with the defect axis named $\hat{\mathbf a}$. For any position $\mathbf r$ on a unit sphere, the new polar angle is defined through $\hat{\mathbf a}\cdot\mathbf r=\cos\theta'$. The associated tangent vector is $\hat{\mathbf e}_{\theta'}=\partial\mathbf r/\partial\theta'=(\cos\theta'\mathbf r-\hat{\mathbf a})/\sin\theta'$. One gets the tangent vector components $c_{\theta}=\hat{\mathbf e}_{\theta'}\cdot\hat{\mathbf e}_{\theta}=-a_{\theta}/\sin\theta'$, $c_{\phi}=\hat{\mathbf e}_{\theta'}\cdot\hat{\mathbf e}_{\phi}=-a_{\phi}/\sin\theta'$ with $a_{\theta}=\hat{\mathbf a}\cdot\hat{\mathbf e}_{\theta}$, $a_{\phi}=\hat{\mathbf a}\cdot\hat{\mathbf e}_{\phi}$ and $a_{\theta}^2+a_{\phi}^2=\sin^2\theta'$. One thus obtains the latitude nematic component along axis $\hat{\mathbf a}$
\begin{align}
Q_{\theta'\theta'}=\hat{\mathbf e}_{\theta'}\cdot\mathbf Q\cdot\hat{\mathbf e}_{\theta'}=Q_{\theta\theta}(c_{\theta}^2-c_{\phi}^2)+2Q_{\theta\phi}c_{\theta}c_{\phi}
\end{align}
Computing the spatial average gives $\langle Q_{\theta\theta}\rangle_{\hat{\mathbf a}}=(1/4\pi)\int\mathrm{d}\phi\mathrm{d}\theta\,\sin\theta Q_{\theta'\theta'}$.

\section{Supplementary figures}
(next pages)
\clearpage

\begin{figure}
\centering
\includegraphics[width=0.9\linewidth]{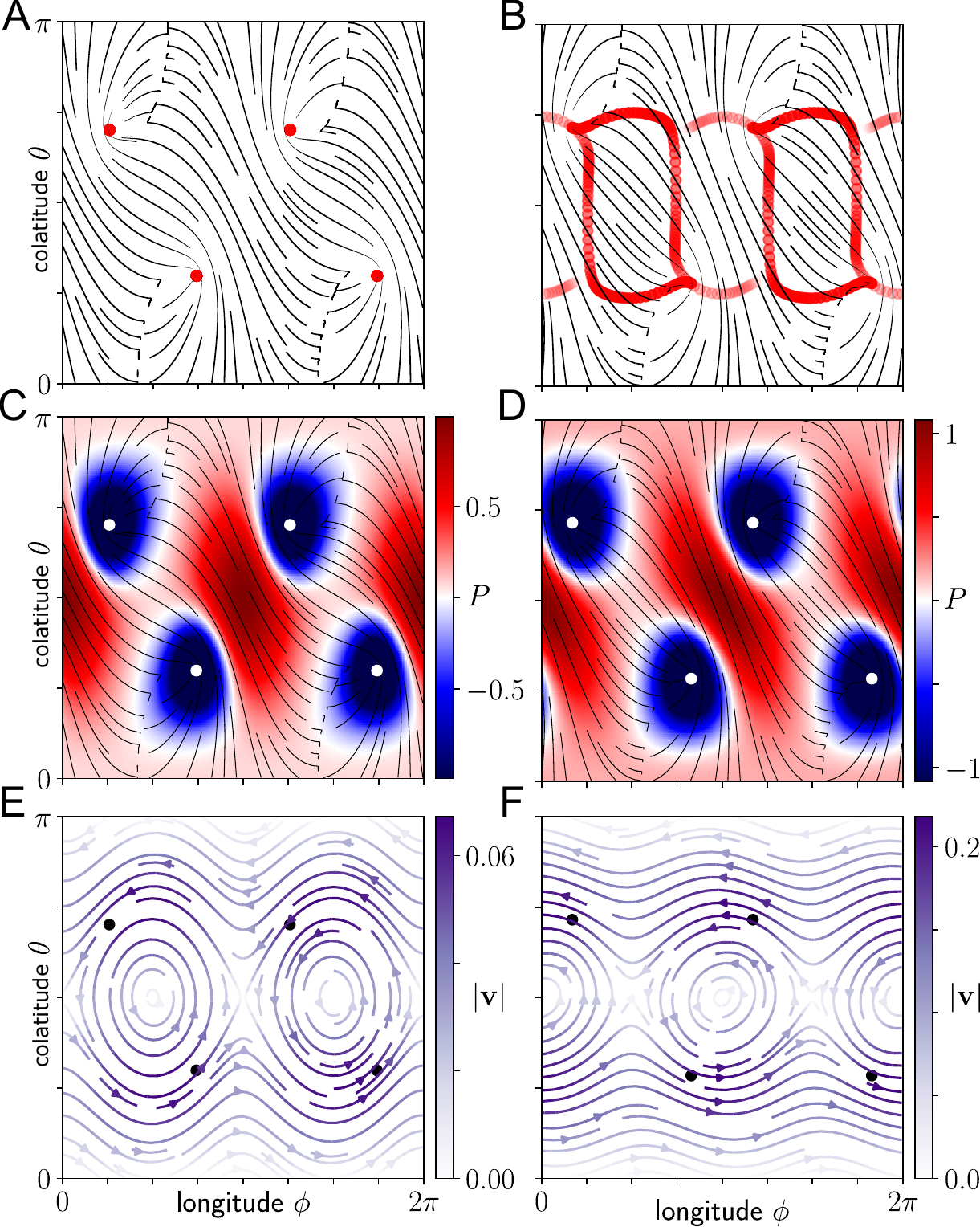}
\caption{\textit{Incompressible surface nematics with active nematic stress $\alpha=-1$.} (A,C,E) Mercator plots of the defect trajectory (A), the pressure field (C) and the velocity field (E) for $\lambda=2$ at $t=20$. (B,D,F) Similar to A,C,E with $\lambda=0$. Other parameters are $\chi=6$, $\xi_s=0.5$, $\beta=0$.}
\label{figS1}
\end{figure}

\begin{figure}
\centering
\includegraphics[width=0.9\linewidth]{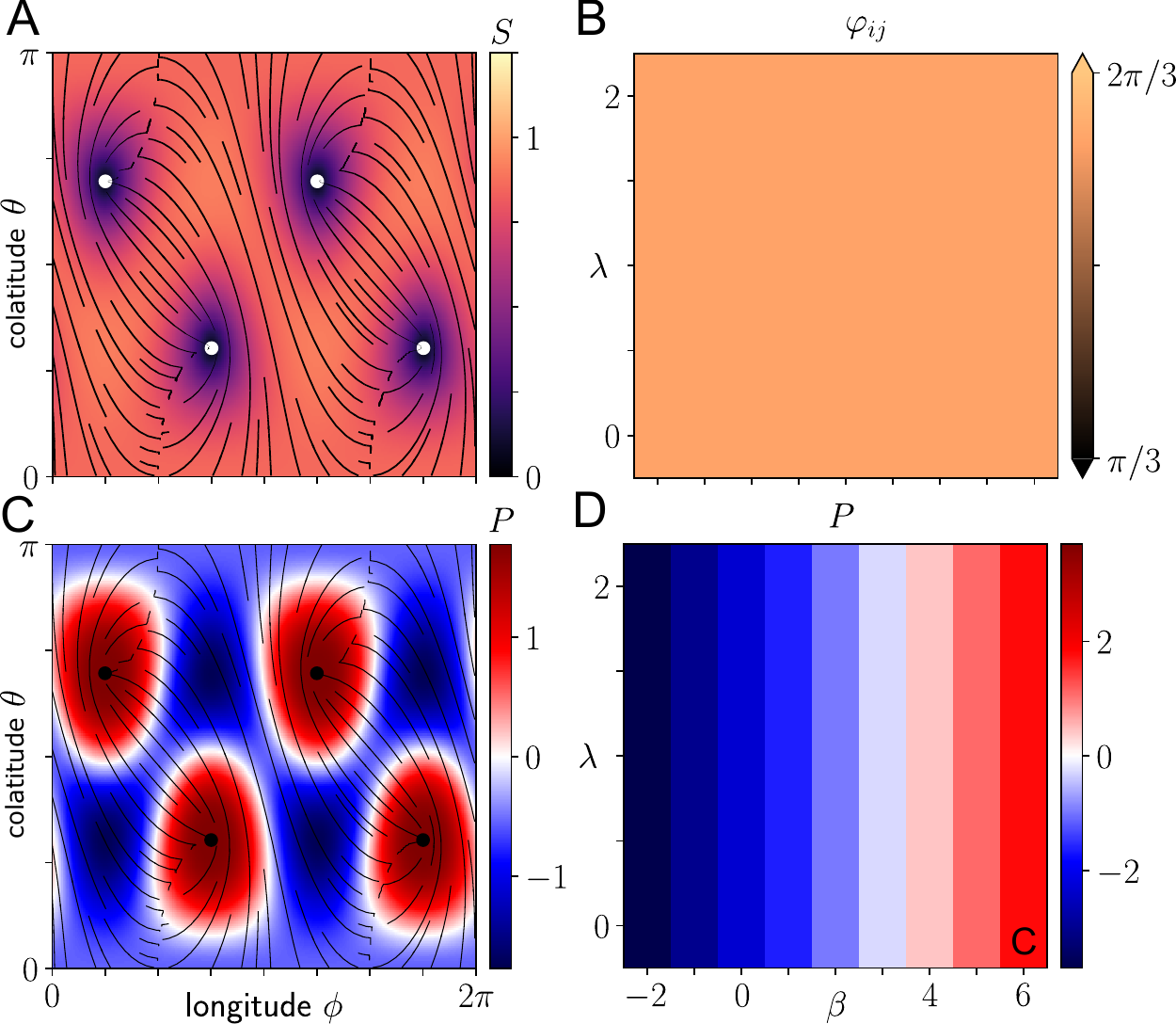}
\caption{\textit{Incompressible surface nematics with active tractions.} (A,C) Mercator plot of the nematic texture (A) and the pressure (C) at $t=5$, for $\beta=6$ and $\lambda=0$.
(B,D) Phase diagrams of the minimal defect pair angular distance $\varphi_{ij}$ (B) and the pressure $P$ at defect positions (D) as a function of the active traction coefficient $\beta$ and the flow-alignment coefficient $\lambda$. Quantities are time-averaged over $t=[2.5,5]$. Other parameters are $\chi=6$, $\xi_s=0.5$, $\alpha=0$.}
\label{figS2}
\end{figure}

\begin{figure}
\centering
\includegraphics[width=0.9\linewidth]{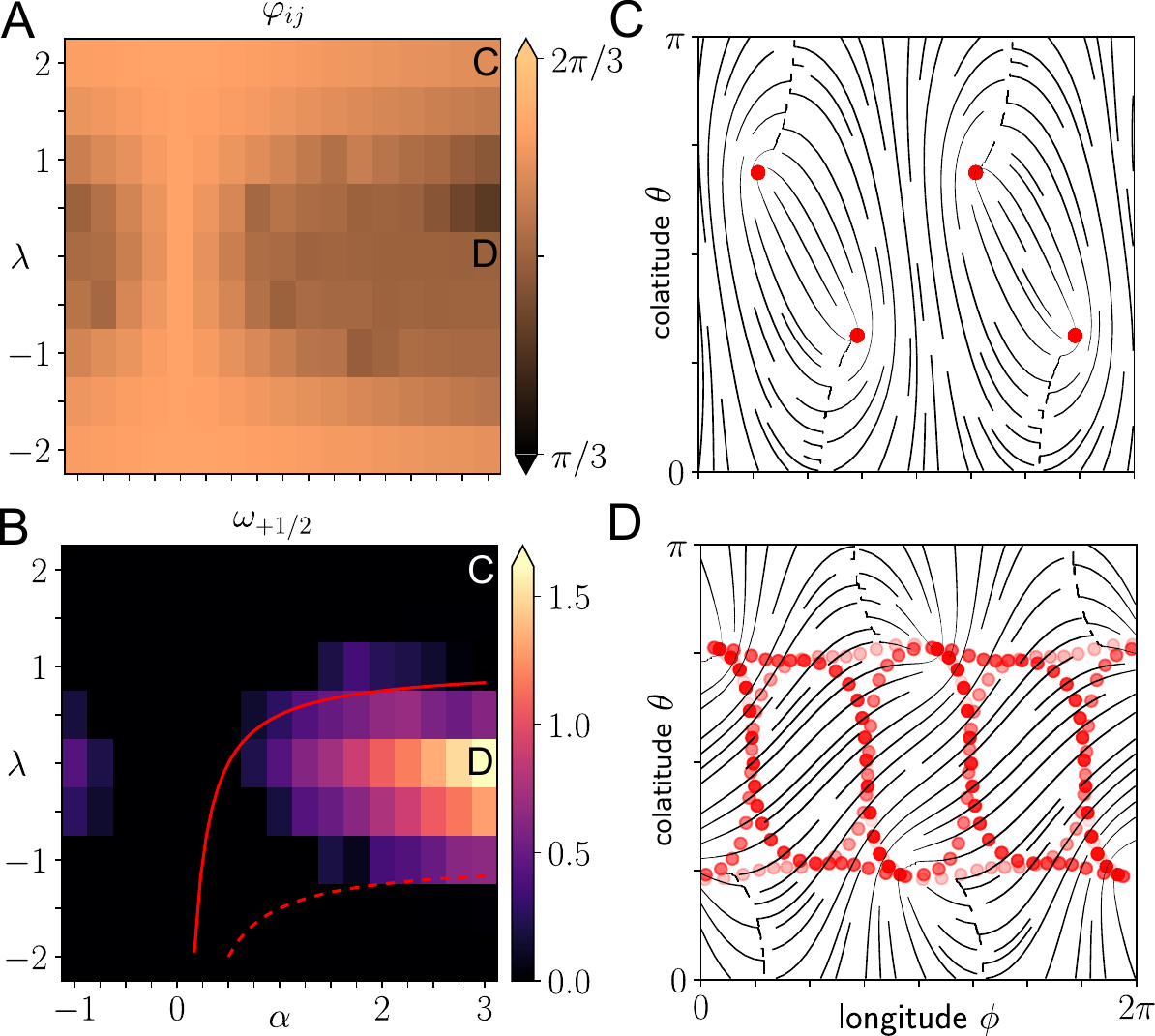}
\caption{\textit{Compressible surface nematics with active nematic stress.}
(A,B) Phase diagrams of the maximum defect pair distance (A) and the mean defect angular velocity (B) as a function of the active nematic stress coefficient $\alpha$ and the flow-alignment coefficient $\lambda$. Quantities are time-averaged over $t=[5,10]$. Red lines on (B) indicate critical lines from the linear stability analysis of a flat nematic fluid, for the splay mode (full) or the bend mode (dashed).
(C,D) Mercator plots of the nematic texture with defect trajectories for $\alpha=3$, at $\lambda=2$ (C) or $\lambda=0$ (D).
Other parameters are $\chi=6$, $\xi_s=0.5$, $\alpha_i=\beta=0$.}
\label{figS3}
\end{figure}

\begin{figure}
\centering
\includegraphics[width=0.9\linewidth]{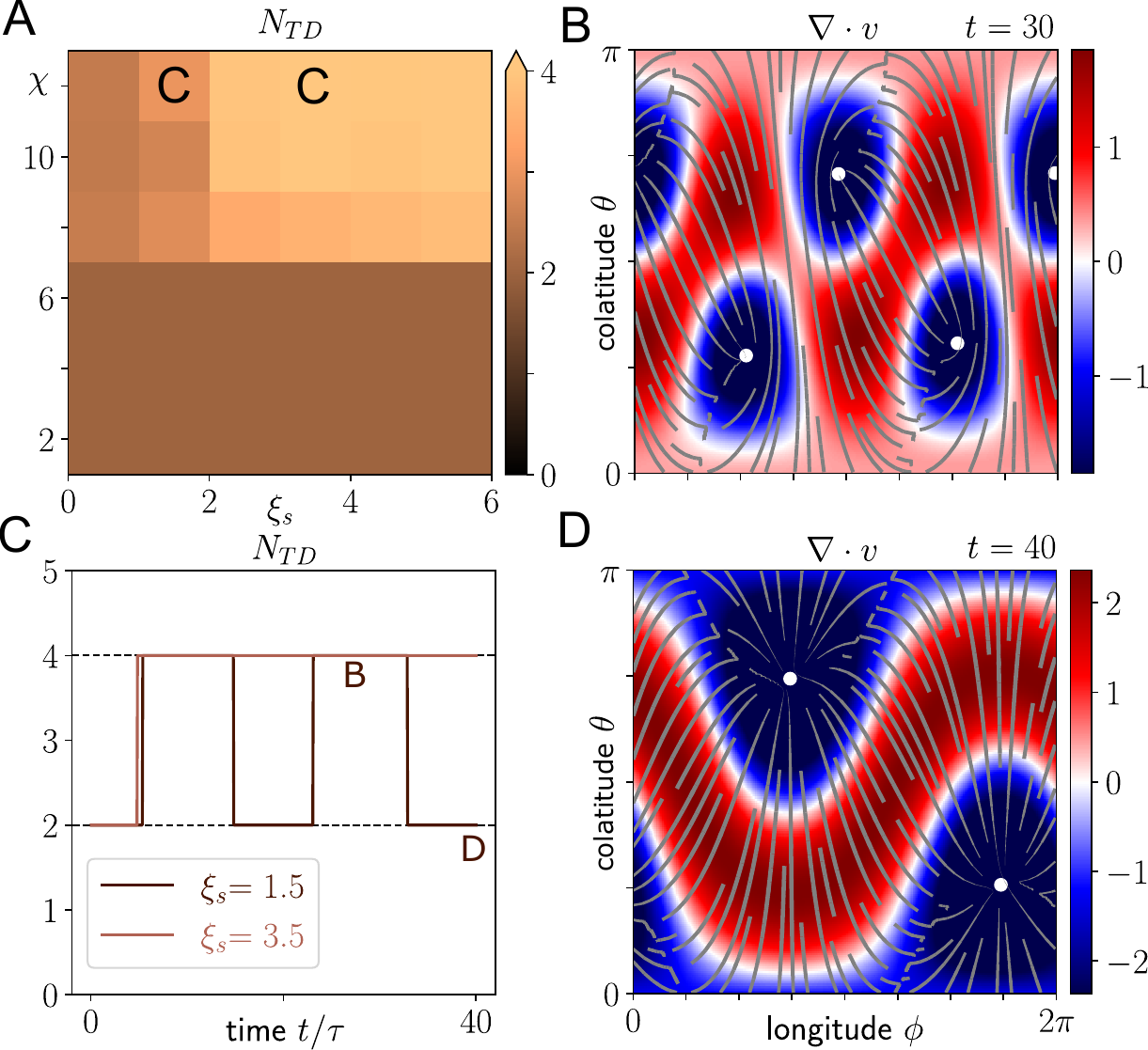}
\caption{\textit{Compressible surface nematics with active tractions and low nematic ordering.}
(A) Phase diagram of the number of defects as a function of the substrate friction $\xi_s$ and the nematic ordering $\chi$ coefficients. Quantities are time-averaged over $t=[5,20]$ for $\chi\leq 6$ and $t=[10,40]$ for $\chi>6$.
(C) Time evolution of the number of defects for $\chi=12$, with $\xi_s=1.5$ (brown) or $\xi_s=3.5$ (light brown).
(B,D) Mercator plots of the nematic texture superposed with flow divergence for $\chi=12$, $\xi_s=1.5$ and $t=30$ (B) or $t=40$ (D).
%(A,B) Phase diagrams of the number of defects (A) and the average latitude nematic $Q_{\theta\theta}$ (B) as a function of the active traction $\beta$ and nematic ordering $\chi$ coefficients. Quantities are time-averaged over $t=[10,20]$.
%(C,D) Mercator plots of the nematic texture at $t=20$, for $\beta=3$ (C) or $\beta=6$ (D) and $\chi=2$.
Other parameters are $\psi_0=0.05\pi$, $\alpha=\alpha_i=0$, $\beta=14$ and $\lambda=1.5$.
}
\label{figS4}
\end{figure}

\clearpage
\newpage

\bibliography{refs}

@article{Dedenon.2023, 
year = {2023}, 
title = {{Density-Polarity Coupling in Confined Active Polar Films: Asters, Spirals, and Biphasic Orientational Phases}}, 
author = {Dedenon, Mathieu and Dessalles, Claire A. and Guillamat, Pau and Roux, Aurélien and Kruse, Karsten and Blanch-Mercader, Carles}, 
journal = {Physical Review Letters}, 
issn = {0031-9007}, 
doi = {10.1103/physrevlett.131.268301}, 
pmid = {38215373}, 
eprint = {2307.15707}, 
pages = {268301}, 
number = {26}, 
volume = {131}
}

@article{Giomi:2013ky, 
year = {2013}, 
title = {{Defect Annihilation and Proliferation in Active Nematics}}, 
author = {Giomi, Luca and Bowick, Mark J and Ma, Xu and Marchetti, MC Cristina}, 
journal = {Phys. Rev. Lett.}, 
issn = {0031-9007}, 
doi = {10.1103/physrevlett.110.228101},
pmid = {23767749}, 
eprint = {1303.4720}, 
url = {http://gateway.webofknowledge.com/gateway/Gateway.cgi?GWVersion=2\&SrcAuth=mekentosj\&SrcApp=Papers\&DestLinkType=FullRecord\&DestApp=WOS\&KeyUT=000319682200023}, 
pages = {228101}, 
number = {22}, 
volume = {110}, 
month = {00}
}

@article{Voituriez:2005jy, 
year = {2005}, 
title = {{Spontaneous flow transition in active polar gels}}, 
author = {Voituriez, R and Joanny, J F and Prost, Jacques}, 
journal = {EPL (Europhysics Letters)}, 
issn = {1286-4854}, 
doi = {10.1209/epl/i2004-10501-2}, 
eprint = {q-bio/0503022}, 
url = {http://stacks.iop.org/0295-5075/70/i=3/a=404?key=crossref.8fe23764d9475a52b853ea2dc49b2914}, 
pages = {404 -- 410}, 
number = {3}, 
volume = {70}, 
month = {01}
}

@article{Chugh2018,
  title = {The actin cortex at a glance},
  volume = {131},
  ISSN = {0021-9533},
  url = {http://dx.doi.org/10.1242/jcs.186254},
  DOI = {10.1242/jcs.186254},
  number = {14},
  journal = {Journal of Cell Science},
  publisher = {The Company of Biologists},
  author = {Chugh,  Priyamvada and Paluch,  Ewa K.},
  year = {2018},
  month = July 
}

@article{Fishkind1993,
  title = {Orientation and three-dimensional organization of actin filaments in dividing cultured cells.},
  volume = {123},
  ISSN = {1540-8140},
  url = {http://dx.doi.org/10.1083/jcb.123.4.837},
  DOI = {10.1083/jcb.123.4.837},
  number = {4},
  journal = {The Journal of cell biology},
  publisher = {Rockefeller University Press},
  author = {Fishkind,  D J and Wang,  Y L},
  year = {1993},
  month = Nov,
  pages = {837–848}
}

@article{Luciano.2024, 
year = {2024}, 
title = {{How multiscale curvature couples forces to cellular functions}}, 
author = {Luciano, Marine and Tomba, Caterina and Roux, Aurélien and Gabriele, Sylvain}, 
journal = {Nature Reviews Physics}, 
doi = {10.1038/s42254-024-00700-9}, 
pages = {246--268}, 
number = {4}, 
volume = {6}
}

@article{Maechler.2019, 
year = {2019}, 
title = {{Curvature-dependent constraints drive remodeling of epithelia}}, 
author = {Maechler, Florian A. and Allier, Cédric and Roux, Aurélien and Tomba, Caterina}, 
journal = {Journal of Cell Science}, 
issn = {0021-9533}, 
doi = {10.1242/jcs.222372}, 
pmid = {30578312}, 
pmcid = {PMC6398478}, 
pages = {jcs222372}, 
number = {4}, 
volume = {132}
}

@article{Doostmohammadi.2018, 
year = {2018}, 
title = {{Active nematics}}, 
author = {Doostmohammadi, Amin and Ignés-Mullol, Jordi and Yeomans, Julia M. and Sagués, Francesc}, 
journal = {Nature Communications}, 
doi = {10.1038/s41467-018-05666-8}, 
pmid = {30131558}, 
pmcid = {PMC6104062}, 
pages = {3246}, 
number = {1}, 
volume = {9}
}

@software{Lanman2020_pyspherical,
  author       = {Adam E. Lanman},
  title        = {pyspherical: Fast Spin-Weighted Spherical Harmonic Transforms},
  year         = {2020},
  version      = {0.0.2},
  publisher    = {GitHub},
  url          = {https://github.com/aelanman/pyspherical},
  note         = {Python package}
}

@article{McEwen2011,
  author  = {Jason D. McEwen and Yves Wiaux},
  title   = {A Novel Sampling Theorem on the Sphere},
  journal = {IEEE Transactions on Signal Processing},
  volume  = {59},
  number  = {12},
  pages   = {5876--5887},
  year    = {2011},
  doi     = {10.1109/TSP.2011.2166394}
}

@article{Trapani2006,
  author  = {Stefano Trapani and Jorge Navaza},
  title   = {Calculation of Spherical Harmonics and Wigner d Functions by FFT. Applications to Fast Rotational Matching in Molecular Replacement and Implementation into AMoRe},
  journal = {Acta Crystallographica Section A},
  volume  = {62},
  number  = {4},
  pages   = {262--269},
  year    = {2006},
  doi     = {10.1107/S0108767306017478}
}

@article{Guillamat:2022, 
year = {2022}, 
title = {{Integer topological defects organize stresses driving tissue morphogenesis}}, 
author = {Guillamat, Pau and Blanch-Mercader, Carles and Pernollet, Guillaume and Kruse, Karsten and Roux, Aurélien}, 
journal = {Nature Materials}, 
issn = {1476-1122}, 
doi = {10.1038/s41563-022-01194-5}, 
pmid = {35145258}, 
pages = {588--597}, 
volume = {21}
}

@article{Turlier:2014hq, 
year = {2014}, 
title = {{Furrow Constriction in Animal Cell Cytokinesis}}, 
author = {Turlier, Herve and Audoly, Basile and Prost, Jacques and Joanny, Jean-Francois}, 
journal = {Biophysical journal}, 
doi = {10.1016/j.bpj.2013.11.014}, 
pmid = {24411243}, 
pmcid = {PMC3907238}, 
url = {http://gateway.webofknowledge.com/gateway/Gateway.cgi?GWVersion=2\&SrcAuth=mekentosj\&SrcApp=Papers\&DestLinkType=FullRecord\&DestApp=WOS\&KeyUT=000329407700016}, 
pages = {114 -- 123}, 
number = {1}, 
volume = {106}
}

@article{Li.2025, 
year = {2025}, 
title = {{Emergence of cellular nematic order is a conserved feature of gastrulation in animal embryos}}, 
author = {Li, Xin and Huebner, Robert J. and Williams, Margot L. K. and Sawyer, Jessica and Peifer, Mark and Wallingford, John B. and Thirumalai, D.}, 
journal = {Nature Communications}, 
doi = {10.1038/s41467-025-61045-0}, 
pmid = {40595575}, 
pmcid = {PMC12217158}, 
pages = {5946}, 
number = {1}, 
volume = {16}
}

@article{Salbreux:2009fp, 
year = {2009}, 
title = {{Hydrodynamics of cellular cortical flows and the formation of contractile rings.}}, 
author = {Salbreux, G and Prost, Jacques and Joanny, J F}, 
journal = {Phys. Rev. Lett.}, 
doi = {10.1103/physrevlett.103.058102}, 
pmid = {19792537}, 
url = {http://eutils.ncbi.nlm.nih.gov/entrez/eutils/elink.fcgi?dbfrom=pubmed\&id=19792537\&retmode=ref\&cmd=prlinks}, 
pages = {058102}, 
number = {5}, 
volume = {103}
}

@article{Copenhagen.2021, 
year = {2021}, 
title = {{Topological defects promote layer formation in Myxococcus xanthus colonies}}, 
author = {Copenhagen, Katherine and Alert, Ricard and Wingreen, Ned S. and Shaevitz, Joshua W.}, 
journal = {Nature Physics}, 
issn = {1745-2473}, 
doi = {10.1038/s41567-020-01056-4}, 
eprint = {2001.03804}, 
pages = {211--215}, 
number = {2}, 
volume = {17}
}

@article{Kawaguchi:2017em, 
year = {2017}, 
title = {{Topological defects control collective dynamics in neural progenitor cell cultures}}, 
author = {Kawaguchi, Kyogo and Kageyama, Ryoichiro and Sano, Masaki}, 
journal = {Nature}, 
issn = {0028-0836}, 
doi = {10.1038/nature22321}, 
pmid = {28403137}, 
url = {http://www.nature.com/articles/nature22321}, 
pages = {327 -- 331}, 
number = {7654}, 
volume = {545}
}

@article{Saw:2017gn, 
year = {2017}, 
title = {{Topological defects in epithelia govern cell death and extrusion}}, 
author = {Saw, Thuan Beng and Doostmohammadi, Amin and Nier, Vincent and Kocgozlu, Leyla and Thampi, Sumesh and Toyama, Yusuke and Marcq, Philippe and Lim, Chwee Teck and Yeomans, Julia M and Ladoux, Benoit}, 
journal = {Nature}, 
issn = {0028-0836}, 
doi = {10.1038/nature21718}, 
pmid = {28406198}, 
url = {http://www.nature.com/articles/nature21718}, 
pages = {212 -- 216}, 
number = {7649}, 
volume = {544}
}

@article{Prost:2015ev, 
year = {2015}, 
title = {{Active gel physics}}, 
author = {Prost, Jacques and Jülicher, Frank and Joanny, J F}, 
journal = {Nature Physics}, 
issn = {1745-2481}, 
doi = {10.1038/nphys3224}, 
url = {http://www.nature.com/doifinder/10.1038/nphys3224}, 
pages = {111 -- 117}, 
number = {2}, 
volume = {11}
}

@article{Vitelli2006,
  title = {Nematic textures in spherical shells},
  volume = {74},
  ISSN = {1550-2376},
  url = {http://dx.doi.org/10.1103/PhysRevE.74.021711},
  DOI = {10.1103/physreve.74.021711},
  number = {2},
  journal = {Physical Review E},
  publisher = {American Physical Society (APS)},
  author = {Vitelli,  V and Nelson,  D. R.},
  year = {2006},
}

@article{Sanchez2012,
  title = {Spontaneous motion in hierarchically assembled active matter},
  volume = {491},
  ISSN = {1476-4687},
  url = {http://dx.doi.org/10.1038/nature11591},
  DOI = {10.1038/nature11591},
  number = {7424},
  journal = {Nature},
  publisher = {Springer Science and Business Media LLC},
  author = {Sanchez,  Tim and Chen,  Daniel T. N. and DeCamp,  Stephen J. and Heymann,  Michael and Dogic,  Zvonimir},
  year = {2012},
  pages = {431–434}
}

@article{Marchetti2013,
  title = {Hydrodynamics of soft active matter},
  volume = {85},
  ISSN = {1539-0756},
  url = {http://dx.doi.org/10.1103/RevModPhys.85.1143},
  DOI = {10.1103/revmodphys.85.1143},
  number = {3},
  journal = {Reviews of Modern Physics},
  publisher = {American Physical Society (APS)},
  author = {Marchetti,  M. C. and Joanny,  J. F. and Ramaswamy,  S. and Liverpool,  T. B. and Prost,  J. and Rao,  Madan and Simha,  R. Aditi},
  year = {2013},
  pages = {1143–1189}
}

@article{Keber2014,
  title = {Topology and dynamics of active nematic vesicles},
  volume = {345},
  ISSN = {1095-9203},
  url = {http://dx.doi.org/10.1126/science.1254784},
  DOI = {10.1126/science.1254784},
  number = {6201},
  journal = {Science},
  publisher = {American Association for the Advancement of Science (AAAS)},
  author = {Keber,  Felix C. and Loiseau,  Etienne and Sanchez,  Tim and DeCamp,  Stephen J. and Giomi,  Luca and Bowick,  Mark J. and Marchetti,  M. Cristina and Dogic,  Zvonimir and Bausch,  Andreas R.},
  year = {2014},
  pages = {1135–1139}
}

@article{Giomi2017,
  title = {Curvature-induced defect unbinding and dynamics in active nematic toroids},
  volume = {14},
  ISSN = {1745-2481},
  url = {http://dx.doi.org/10.1038/nphys4276},
  DOI = {10.1038/nphys4276},
  number = {1},
  journal = {Nature Physics},
  publisher = {Springer Science and Business Media LLC},
  author = {Ellis,  Perry W. and Pearce,  Daniel J. G. and Chang,  Ya-Wen and Goldsztein,  Guillermo and Giomi,  Luca and Fernandez-Nieves,  Alberto},
  year = {2017},
  pages = {85–90}
}

@article{Sano2017,
	title = {Topological defects control collective dynamics in neural progenitor cell cultures},
	volume = {545},
	copyright = {2017 Macmillan Publishers Limited, part of Springer Nature. All rights reserved.},
	issn = {1476-4687},
	url = {https://www.nature.com/articles/nature22321},
	doi = {10.1038/nature22321},
	number = {7654},
	urldate = {2023-05-20},
	journal = {Nature},
	author = {Kawaguchi, Kyogo and Kageyama, Ryoichiro and Sano, Masaki},
	year = {2017},
	pages = {327--331},
}

@article{Duclos2018,
  title = {Spontaneous shear flow in confined cellular nematics},
  volume = {14},
  ISSN = {1745-2481},
  url = {http://dx.doi.org/10.1038/s41567-018-0099-7},
  DOI = {10.1038/s41567-018-0099-7},
  number = {7},
  journal = {Nature Physics},
  publisher = {Springer Science and Business Media LLC},
  author = {Duclos,  G. and Blanch-Mercader,  C. and Yashunsky,  V. and Salbreux,  G. and Joanny,  J.-F. and Prost,  J. and Silberzan,  P.},
  year = {2018},
  pages = {728–732}
}

@article{Pearce2019,
  title = {Geometrical Control of Active Turbulence in Curved Topographies},
  volume = {122},
  ISSN = {1079-7114},
  url = {http://dx.doi.org/10.1103/PhysRevLett.122.168002},
  DOI = {10.1103/physrevlett.122.168002},
  number = {16},
  journal = {Physical Review Letters},
  publisher = {American Physical Society (APS)},
  author = {Pearce,  D.J.G. and Ellis,  Perry W. and Fernandez-Nieves,  Alberto and Giomi,  L.},
  year = {2019} 
}

@article{Ehrig2019,
  title = {Surface tension determines tissue shape and growth kinetics},
  volume = {5},
  ISSN = {2375-2548},
  url = {http://dx.doi.org/10.1126/sciadv.aav9394},
  DOI = {10.1126/sciadv.aav9394},
  number = {9},
  journal = {Science Advances},
  publisher = {American Association for the Advancement of Science (AAAS)},
  author = {Ehrig,  S. and Schamberger,  B. and Bidan,  C. M. and West,  A. and Jacobi,  C. and Lam,  K. and Kollmannsberger,  P. and Petersen,  A. and Tomancak,  P. and Kommareddy,  K. and Fischer,  F. D. and Fratzl,  P. and Dunlop,  John W. C.},
  year = {2019}
}

@article{Vasil2019,
  title = {Tensor calculus in spherical coordinates using Jacobi polynomials. Part-I: Mathematical analysis and derivations},
  volume = {3},
  ISSN = {2590-0552},
  url = {http://dx.doi.org/10.1016/j.jcpx.2019.100013},
  DOI = {10.1016/j.jcpx.2019.100013},
  journal = {Journal of Computational Physics: X},
  publisher = {Elsevier BV},
  author = {Vasil,  Geoffrey M. and Lecoanet,  Daniel and Burns,  Keaton J. and Oishi,  Jeffrey S. and Brown,  Benjamin P.},
  year = {2019},
  month = June,
  pages = {100013}
}

@article{MaroudasSacks2020,
  title = {Topological defects in the nematic order of actin fibres as organization centres of Hydra morphogenesis},
  volume = {17},
  ISSN = {1745-2481},
  url = {http://dx.doi.org/10.1038/s41567-020-01083-1},
  DOI = {10.1038/s41567-020-01083-1},
  number = {2},
  journal = {Nature Physics},
  publisher = {Springer Science and Business Media LLC},
  author = {Maroudas-Sacks,  Yonit and Garion,  Liora and Shani-Zerbib,  Lital and Livshits,  Anton and Braun,  Erez and Keren,  Kinneret},
  year = {2020},
  pages = {251–259}
}

@article{Vafa2022,
  title = {Active Nematic Defects and Epithelial Morphogenesis},
  volume = {129},
  ISSN = {1079-7114},
  url = {http://dx.doi.org/10.1103/PhysRevLett.129.098102},
  DOI = {10.1103/physrevlett.129.098102},
  number = {9},
  journal = {Physical Review Letters},
  publisher = {American Physical Society (APS)},
  author = {Vafa,  Farzan and Mahadevan,  L.},
  year = {2022} 
}

@article{Bouck2023,
  title = {A Hydrodynamical Model of Nematic Liquid Crystal Films with a General State of Orientational Order},
  volume = {34},
  ISSN = {1432-1467},
  url = {http://dx.doi.org/10.1007/s00332-023-09970-6},
  DOI = {10.1007/s00332-023-09970-6},
  number = {1},
  journal = {Journal of Nonlinear Science},
  publisher = {Springer Science and Business Media LLC},
  author = {Bouck,  Lucas and Nochetto,  Ricardo H. and Yushutin,  Vladimir},
  year = {2023},
  month = Oct 
}

@article{Dessalles2025,
  title = {Interplay of actin nematodynamics and anisotropic tension controls endothelial mechanics},
  volume = {21},
  ISSN = {1745-2481},
  url = {http://dx.doi.org/10.1038/s41567-025-02847-3},
  DOI = {10.1038/s41567-025-02847-3},
  number = {6},
  journal = {Nature Physics},
  publisher = {Springer Science and Business Media LLC},
  author = {Dessalles,  Claire A. and Cuny,  Nicolas and Boutillon,  Arthur and Salipante,  Paul F. and Babataheri,  Avin and Barakat,  Abdul I. and Salbreux,  Guillaume},
  year = {2025},
  pages = {999–1008}
}

@article{Ravichandran2025,
  title = {Topology changes of
            Hydra
            define actin orientation defects as organizers of morphogenesis},
  volume = {11},
  ISSN = {2375-2548},
  url = {http://dx.doi.org/10.1126/sciadv.adr9855},
  DOI = {10.1126/sciadv.adr9855},
  number = {3},
  journal = {Science Advances},
  publisher = {American Association for the Advancement of Science (AAAS)},
  author = {Ravichandran,  Yamini and Vogg,  Matthias and Kruse,  Karsten and Pearce,  Daniel J. G. and Roux,  Aurélien},
  year = {2025}
}

@article{Zhao2025,
	title = {Integer topological defects offer a methodology to quantify and classify active cell monolayers},
	volume = {16},
	issn = {2041-1723},
	url = {https://www.nature.com/articles/s41467-025-57783-w},
	doi = {10.1038/s41467-025-57783-w},
	number = {1},
	urldate = {2026-03-30},
	journal = {Nature Communications},
	author = {Zhao, Zihui and Li, He and Yao, Yisong and Zhao, Yongfeng and Serra, Francesca and Kawaguchi, Kyogo and Zhang, Hepeng and Sano, Masaki},
	month = mar,
	year = {2025},
	pages = {2452},
}

@article{Guillamat2026,
  title = {Guidance of cellular nematic elastomers into shape-programmable living surfaces},
  volume = {392},
  ISSN = {1095-9203},
  url = {http://dx.doi.org/10.1126/science.adz9174},
  DOI = {10.1126/science.adz9174},
  number = {6795},
  journal = {Science},
  publisher = {American Association for the Advancement of Science (AAAS)},
  author = {Guillamat,  Pau and Mirza,  Waleed and Bal,  Pradeep K. and Gómez-González,  Manuel and Roca-Cusachs,  Pere and Arroyo,  Marino and Trepat,  Xavier},
  year = {2026},
  pages = {317–323}
}

@article{Rembert2026,
  title = {Nematic order in cellular tissues: a standardized framework and anomalous defect dynamics},
  url = {http://dx.doi.org/10.64898/2026.04.22.719598},
  DOI = {10.64898/2026.04.22.719598},
  publisher = {openRxiv},
  journal = {bioRxiv},
  author = {Rembert,  Nicolas and Dedenon,  Mathieu and Roux,  Aurélien and Dessalles,  Claire A.},
  year = {2026}
}

%\clearpage

\end{document}